\pdfoutput=1

\documentclass[11pt]{article}

\usepackage{ACL2023}

\usepackage{times}
\usepackage{latexsym}
\usepackage{graphicx}
\usepackage{fancybox}
\usepackage{amsmath}
\usepackage{amssymb}
\usepackage{booktabs}
\usepackage{multirow}
\usepackage{booktabs} 

\usepackage[T1]{fontenc}

\usepackage[utf8]{inputenc}

\usepackage{microtype}

\usepackage{inconsolata}

\usepackage{stmaryrd}
\usepackage{amsfonts}
\usepackage{amssymb}
\usepackage{amsmath}
\usepackage{bbm}
\usepackage{xspace}
\usepackage{todonotes}
\usepackage[linesnumbered,ruled,vlined]{algorithm2e}
\usepackage{tabularx,booktabs,multirow}  
\usepackage{subcaption}
\usepackage{enumerate}
\usepackage{siunitx}
\usepackage{amsthm}

\usepackage{amsmath,amsfonts,bm}

\def\eqref#1{equation~\ref{#1}}

\def\1{\bm{1}}

\def\ve{{\bm{e}}}

\def\vm{{\bm{m}}}

\def\mE{{\bm{E}}}

\def\mH{{\bm{H}}}

\def\mT{{\bm{T}}}

\def\mW{{\bm{W}}}
\def\mX{{\bm{X}}}

\DeclareMathAlphabet{\mathsfit}{\encodingdefault}{\sfdefault}{m}{sl}
\SetMathAlphabet{\mathsfit}{bold}{\encodingdefault}{\sfdefault}{bx}{n}

\def\gM{{\mathcal{M}}}

\def\gO{{\mathcal{O}}}

\def\gX{{\mathcal{X}}}

\theoremstyle{definition}
\newtheorem{definition}{Definition}[section]

\theoremstyle{remark}

\newcommand{\ie}{\text{i.e.}\xspace}
\newcommand{\eg}{\text{e.g.}\xspace}

\newcommand{\fscore}{$F_1$-score\xspace}

\newcommand{\ourmethod}{ALGEN}

\title{ ALGEN: Few-shot Inversion Attacks on Textual Embeddings \\using Alignment and Generation}

\author{
  Yiyi Chen\textsuperscript{1},\space
  Qiongkai Xu\textsuperscript{2}\thanks{\ \ Corresponding author.},\space
  Johannes Bjerva\textsuperscript{1} \\
  \textsuperscript{1}Department of Computer Science, Aalborg University, Copenhagen, Denmark \\
  \textsuperscript{2}School of Computing, FSE, Macquarie University, Sydney, Australia \\
  \texttt{yiyic@cs.aau.dk, qiongkai.xu@mq.edu.au, jbjerva@cs.aau.dk}
}

\begin{document}

\maketitle
\begin{abstract}
With the growing popularity of Large Language Models (LLMs) and vector databases, private textual data is increasingly processed and stored as numerical embeddings.
However, recent studies have proven that such embeddings are vulnerable to inversion attacks, where original text is reconstructed to reveal sensitive information.
Previous research has largely assumed access to millions of sentences to train attack models, e.g., through data leakage or nearly unrestricted API access.
With our method, \emph{a single data point} is sufficient for a partially successful inversion attack.
With as little as 1k data samples, performance reaches an optimum across a range of black-box encoders, without training on leaked data.
We present a Few-shot Textual Embedding Inversion Attack using \textbf{AL}ignment and \textbf{GEN}eration (\textbf{\ourmethod}), 
by aligning victim embeddings to the attack space and using a generative model to reconstruct text.
We find that \textbf{\ourmethod} attacks can be effectively transferred across domains and languages, revealing key information.
We further examine a variety of defense mechanisms against \textbf{\ourmethod}, and find that none are effective, highlighting the vulnerabilities posed by inversion attacks.
By significantly lowering the cost of inversion and proving that embedding spaces can be aligned through one-step optimization, we establish a new textual embedding inversion paradigm with broader applications for embedding alignment in NLP.\footnote{We open-source our code \url{https://github.com/siebeniris/ALGEN}.}



\end{abstract}

\begin{figure}[t!]
    \centering
    \includegraphics[width=\linewidth]{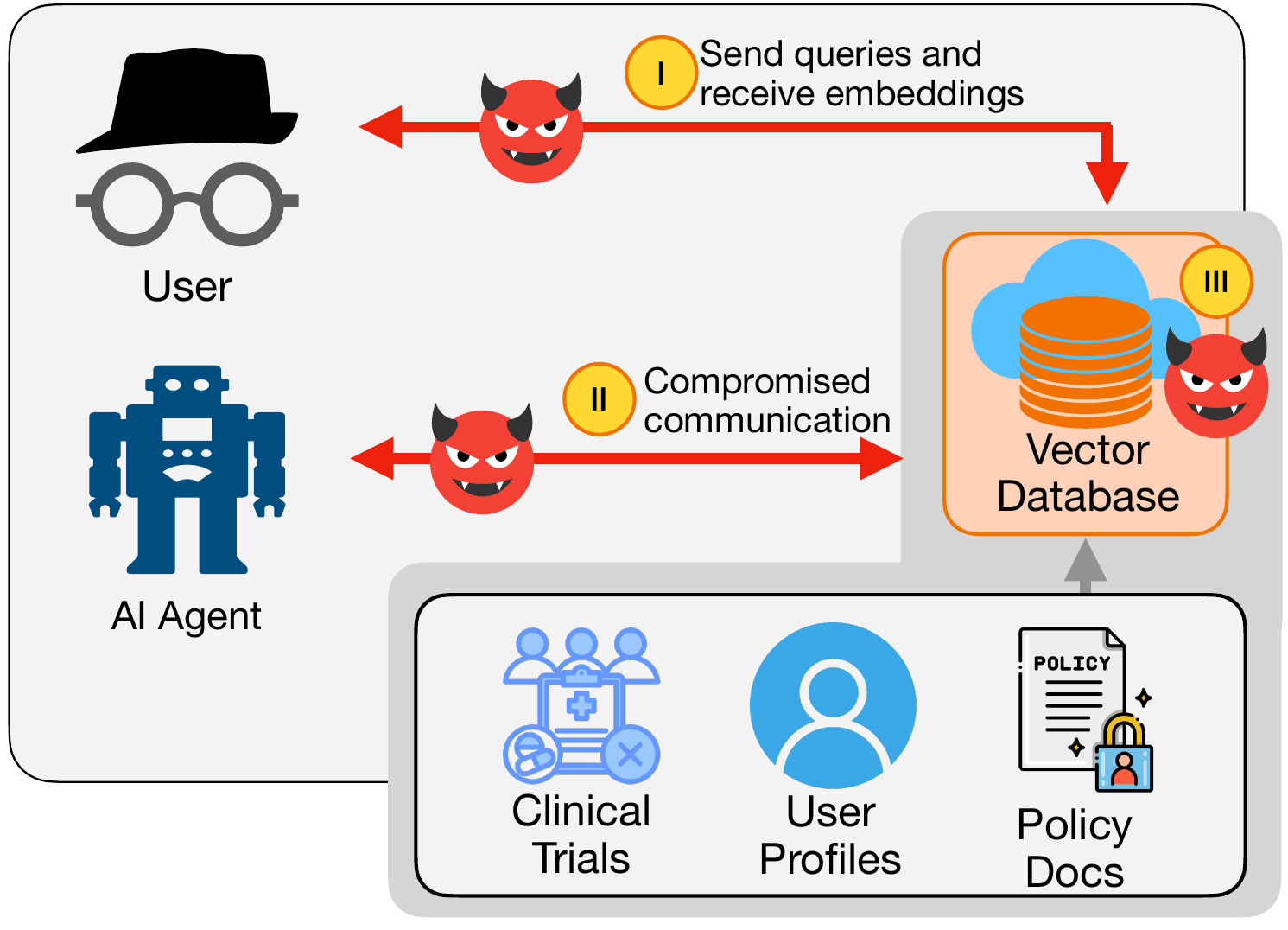}
    \caption{An illustration of inversion attacks on textual embeddings stored in a vector DB, in scenarios where (I) a user exploits API access to extract excessive embeddings to train attack model; (II) a generative AI agent's interaction channel with the DB is compromised; (III) the DB is misconfigured by an insider to expose private embeddings. }
    \label{fig:schema}
\end{figure}

\section{Introduction}
Large Language Models (LLMs) such as OpenAI's GPT series~\citep{radford2018improving, radford2019language,brown2020languagemodelsfewshotlearners,openai2024gpt4technicalreport} and Claude from Anthropic,\footnote{\url{https://www.anthropic.com/claude}}  have become essential across a wide range of applications, extending far beyond natural language processing (NLP). 
These models are deeply integrated into people's daily lives and business operations, e.g., powering search engines, virtual assistants, and content generation. 
A critical component enabling the efficiency of these applications is vector databases (DB), which allow for fast and scalable retrieval and processing of high-dimensional vector representations. 
Companies such as Pinecone and Weaviate,
provide vector DB services and build AI services on top of them.\footnote{\url{https://www.pinecone.io}, \url{https://weaviate.io}}
In a recent Google whitepaper on generative AI agents~\citep{agents_google},
vector DBs are considered one of the essential components enabling such agents through external sources. 
Retrieval-augmented generation (RAG) is another common use case in leveraging vector DBs to generate more diverse and factually grounded responses~\citep{rag_lewis_2020}.

While applications such as these benefit from vector DBs, the potential security and privacy risks permeate the process.
Fig.~\ref{fig:schema} illustrates three separate threat scenarios where a vector DB can be exploited to expose private and sensitive information:
(I) a malicious user exploits the model API to extract  embeddings to train an attack model;
(II) when an AI agent interacts with the DB, a malicious attacker can compromise the communication channel to intercept sensitive data;
(III) a misconfigured vector DB may expose private data, either through access vulnerabilities or insider threats.
%
The attacker can train an attack model (\eg an embedding-to-text generator) to reconstruct text from intercepted embeddings, which might contain sensitive, private, or proprietary information. 
This so-called \textit{embedding inversion attack} poses significant risks and potential harm.


Previous work has demonstrated the feasibility and detrimental effects of inversion attacks~\citep{10.1145/3372297.3417270, li-etal-2023-sentence}. 
However, either a massive amount of intercepted (victim) embeddings and their texts are required for training an attack model~\citep{ morris2023text}, or the attack is conducted under white-box settings, where the model parameters and architecture are known to the attacker~\citep{10.1145/3372297.3417270}. 
Moreover, inversion attacks have been demonstrated to threaten multiple languages, especially lower-resource ones~\citep{chen2024typ,chen2024text}.

We propose a Few-shot Textual Embedding Inversion Attack using \textbf{AL}ignment and \textbf{GEN}eration (\textbf{\ourmethod}), 
to first align victim embeddings to the attack embedding space, and then reconstruct text from the aligned embeddings using the generative attack model.
In contrast to previous work, we investigate inversion attacks using a small handful of samples -- e.g., a Rouge-L score of 10 can be reached by leveraging \emph{a single leaked data point}. 
Our work makes the following main contributions:

\begin{itemize}
\item We propose and verify the effectiveness of a novel few-shot inversion attack, which drastically reduces the cost and complexity of such attacks, making them plausible real-world threats. 

\item We demonstrate the transferability of the inversion attack across various languages, models and domains.

\item We examine several established defense mechanisms, none of which are successful mitigation strategies for this attack, highlighting the new security and privacy vulnerabilities of embeddings in vector databases.

\end{itemize}

\section{Related Work}

\subsection{Textual Embedding Inversion Attacks}
Textual embedding inversion attack aims to learn the inversion function that reconstructs the original textual inputs given their embeddings. 
\citet{10.1145/3372297.3417270} demonstrates that it is possible to recover over half of the input words from a text embedding without preserving their order.
\citet{li-etal-2023-sentence} starts to treat the inversion attacks as a generation task, generating coherent and contextually similar sentences compared to the original text.
~\citet{morris2023text} adopts an iterative approach to train the attack model by parameterizing attack and hypothesis embeddings based on decoded text from the previous step, which results in exact matches between original and reconstructed text in certain settings. 
~\citet{huang_transferable_2024} implements adversarial training to align victim embeddings to attack embeddings, making them not differentiable. 
~\citet{chen2024typ,chen2024text} expand inversion attacks beyond English embeddings to multilingual spaces, leveraging linguistic typology to investigate inversion attack performance, finding that certain languages are particularly vulnerable.

However, all existing works in embedding inversion attacks require an enormous amount of data leakage to train the generative attack models, such as 100k samples for~\citet{li-etal-2023-sentence}, 1-5 million for~\citet{morris2023text, chen2024typ,chen2024text} and 8k for~\citet{huang_transferable_2024}.
In comparison, our proposed approach \textbf{\ourmethod} does not require training an attack model on leaked/private embeddings, and the inversion attack succeeds with few leaked data, we additionally experiment on multiple languages.


\subsection{Embedding Alignment}
Embedding alignment has continuously progressed in NLP with the development of embedding representations and LLMs.
In the early stages, a common approach involved independently training monolingual word vectors~\citep{10.5555/2999792.2999959} and then learning a mapping between source and target language embeddings using a bilingual dictionary~\citep{mikolov2013exploiting,smith2017offline,artetxe-etal-2017-learning}. 
When this mapping is restricted to an orthogonal linear transformation, the optimal word pair alignment can be computed in closed form~\citep{artetxe-etal-2016-learning,schonemann1966generalized}. 
In contrast,~\citet{lample2018word} introduce an unsupervised method for aligning word embedding spaces, incorporating cross-domain similarity adaptation to address the hubness problem.

With the advancement of contextualized embeddings since the emergence of LLMs such as BERT~\citep{devlin2018bert}, the focus shifted to the alignment of contextual word representations~\citep{schuster-etal-2019-cross, aldarmaki-diab-2019-context,wang-etal-2019-cross,alqahtani-etal-2021-using-optimal,cao2020multilingualalignmentcontextualword, jalili-sabet-etal-2020-simalign}.
Moreover, sentence embedding alignment has been operated in lifelong relation extraction with a linear transformation~\citep{wang-etal-2019-sentence}, aligning encoders in different languages to evaluate crosslingual transfer~\citep{conneau2018xnlievaluatingcrosslingualsentence},  building parallel data for machine translation~\citep{krahn2023sentenceembeddingmodelsancient}. 
In comparison, \textbf{\ourmethod} aligns sentence embeddings from different models to conduct embedding inversion attacks, but it can also be applied in embedding alignment in general.

\subsection{Mitigating Embedding Inversion Attacks}
Most research on textual embedding inversion focuses on attacks~\cite {li-etal-2023-sentence, huang_transferable_2024,chen2024typ}.
While~\citet{10.1145/3372297.3417270} adopt an adversarial training approach to mitigate the risks of inversion attacks, this method is ineffective for defending textual embeddings in black-box settings.
To defend against inversion attacks while maintaining embedding utility in downstream tasks, \citet{morris2023text} propose inserting Gaussian noise as a defense mechanism.
Expanding inversion attacks into multilingual space using the same method,~\citet{chen2024text} find that Gaussian noise effectively protects monolingual embeddings but is less effective for multilingual ones.

Differential privacy (DP) limits the impact of individual element~\citep{dwork2014algorithmic}, and has been shown to preserve the privacy of the extracted representation from text, when applied during model training~\citep{lyu-etal-2020-differentially}.
To ensure sequence-level metric-based local DP, which can be employed during inference, a sentence embedding sanitization pipeline has been developed, maintaining non-private task accuracy and effectively thwarting privacy threats of membership inference attacks~\citep{du2023sanitizing}.
Pertinent to vector DBs, Watermarking EaaS with Linear Transformation (WET) introduces a method that applies linear transformations to embeddings to implant watermarks to counter paraphrasing vulnerabilities~\citep{shetty2024wet}.

In this work, we examine these defenses against \textbf{\ourmethod}, and discuss the potentials and challenges of defending embeddings from inversion attacks.

\section{Methodology}

\begin{figure}[t!]
    \centering
    \includegraphics[width=\linewidth]{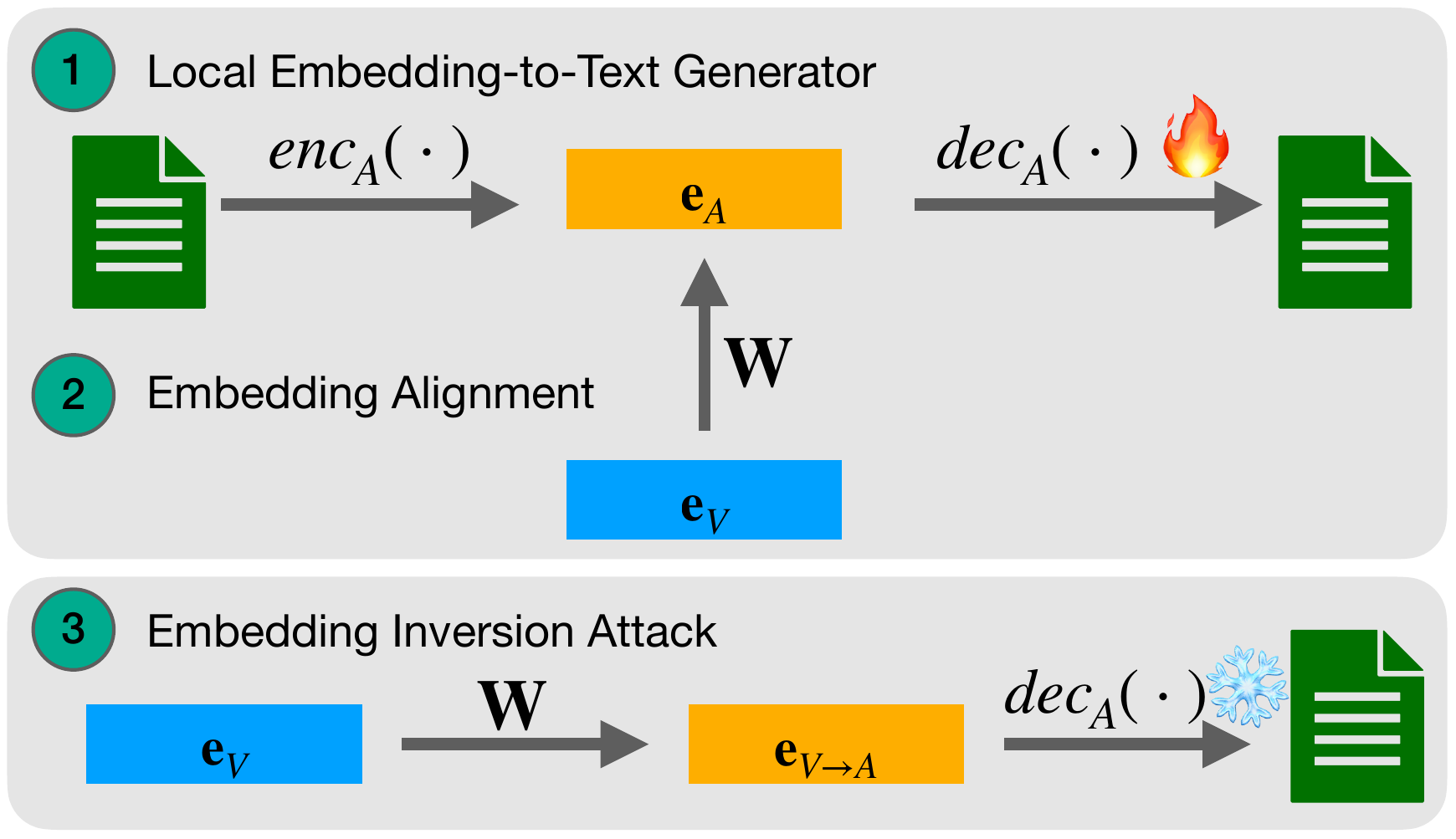}
    \caption{Three steps for Few-shot Inversion Attack, (1) Train a Local Embedding-to-Text Generation Model; (2) Transform \textcolor{blue}{victim embeddings $\ve_{V}$} to the \textcolor{orange}{attack embeddings space $A$} with matrix $\mW$; and (3) Textual embedding inversion attack. 
    }
    \label{fig:method}
\end{figure}



We explore a situation in which a malicious attacker gains access to a \emph{limited set of embeddings}, and attempts to reconstruct private and sensitive text data. 
We propose \textbf{\ourmethod} to circumvent the disadvantage of scarce data and leverage a pretrained encoder-decoder to align the victim embeddings to the attack space and reconstruct the texts.
We note the victim and attack embedding spaces as $V$ and $A$, respectively.
As illustrated in Fig.~\ref{fig:method}, the framework consists of three steps: 
(1) we train an embedding-to-sequence generation model by fine-tuning a pre-trained decoder~$dec_{A}(\cdot)$; (2) we align the embeddings from a black-box victim encoder $\ve_V$  
to attack embeddings in attacker's model space $\ve_A$;
(3) the attacker leverages the capability of the generation model $dec_{A}(\cdot)$ with the embedding alignment model $\mW$ to reconstruct the original text from $\ve_{V}$ to $\ve_{V \rightarrow A}$ and finally to text.

\subsection{Local Embedding-to-Text Generator}
To train the local embedding-to-text generator, we use a publicly available text corpus, noted as $D_L$. 
Given a sentence $x\in D_{L} $, and attack encoder $enc_{A}(\cdot)$, the token embeddings are obtained $\mH=enc_{A}(x)\in \mathbb{R}^{s\times n}$
where $s$ is the sequence length and $n$ the embedding dimension. 
The sentence embeddings are computed through mean pooling the last hidden embeddings $\ve_{A}= \sum^{s}_{j=1}\vm_j \mH_j/ \sum^{s}_{j=1}\vm_j \in \mathbb{R}^{n}$, where $\vm\in \{0,1\}^{s}$ is the attention mask for the sequence $x$.
Furthermore, L2 normalization is implemented on the sentence embeddings, i.e., ${\ve_{A}}/{\|\ve_{A}\|}$
as sentence embedding normalization has proven beneficial in avoiding overfitting and inducing faster convergence in fine-tuning~\citep{aboagye2022normalization}.
The pre-trained decoder $dec_{A}(\cdot)$, parameterized by $\theta$, processes the embeddings to produce an output sequence $\hat{x}=(\hat{x}_1, \hat{x}_{2},\dots, \hat{x}_{s})$, where $\hat{x}_i$ are tokens in the predicted output sequence. We train the embedding-to-text generator by fine-tuning $dec_{A}(\cdot)$, by minimizing the cross-entropy loss:
\begin{equation}
    \mathcal{L}(\theta)= - \sum_{i=1}^{s} \log \mathbb{P}(\hat{x}_i | \hat{x}_{<i}, \ve_{A}; \theta),
\end{equation}
where $\mathbb{P}(\hat{x}_i | \hat{x}_{<i}, \ve_{A}; \theta)$ is the probability of predicting token $\hat{x}_i$ given the previous tokens $\hat{x}_{<i}$ and the input sentence embeddings $\ve_{A}$. 

Notably, attackers can easily retrieve a large corpus $D_L$ to train this generator model, as it operates independently of any victim models.
The challenge of aligning victim semantics to target embedding space is addressed in the next subsection.


\subsection{Embedding Space Alignment}
Suppose there is a leaked data pair $(\mX, \mE_V)$,  given that $\mX \subseteq D_{V}$ is the victim dataset with $b \in \mathbb{N}$ samples, and embedding matrix $\mE_V=enc_{V}(\mathbf{\mX})$, where $enc_{V}$ is the black-box viticm encoder. 
To align the victim embeddings to the attack space $A$, we obtain the embedding matrix $\mE_{A}=enc_{A}(\mathbf{\mX})$ given the leaked text dataset, and seek a solution to solve the system 
\begin{equation}
    \mE_{V}\mW \approx \mE_{A},
\end{equation}
and the best possible $\mW \in \mathbb{R}^{m\times n}$, given $\mE_{V}\in \mathbb{R}^{b\times m}$ and $\mE_{A}\in \mathbb{R}^{b\times n}$, where $n$ and $m$ are the regarding embedding dimensions of victim and attacker embeddings. 
While there is no exact solution to the system, our approach is to minimize their deviation, $e=\mE_{A}-\mE_{V}\mW$. 
Taking the square of the error by each sample, the objective is to minimize the following:
\begin{equation}
    \min_{\mW}
    \sum_{i=1}^{b}
    \|\ve^{i}_{A}- \ve^{i}_{V} \cdot \mW\|^{2}.
\end{equation}
The solution to this least squares loss is:
\begin{equation}
\mW = (\mE^{T}_V \mE_{V})^{-1} \mE_{V}^{T} \mE_{A},
\end{equation}
where $(\mE^{T}_V \mE_V)^{-1} \mE_{V}^{T}$ is the \textit{Moore-Penrose Inverse} of $\mE_V$ (see the detailed derivation in Appendix~\ref{normal_equation}).

The aligned embedding from $V$ to $A$ is thus:
\begin{equation}
    \mE_{V \rightarrow A} = \mE_{V}  \mW,
\end{equation}
where $\mE_{V \rightarrow A}\in \mathbb{R}^{b\times n} $. Implementing this alignment does not require any training; it is a one-step linear scaling. 
Moreover, aligning using $D_{V}$ with a batch size $b$ varying from 1 to 1,000, we observe that even with as few as 30 samples the Rouge-L score exceeds 20, and a reasonably successful attack can be initiated with only a single data point.
The density distribution of the alignment transformation matrix $\mW$'s weights of encoders remains consistent across different datasets (see Fig~\ref{fig:W_alignment_density}).

\subsection{Textual Embedding Inversion Attack}
Given the attack model, \ie the local embedding-to-text generator $dec_{A}(\cdot)$ and the $\ve_V$ to $\ve_A$ alignment model, 
and a body of eavesdropped embeddings $\mE_V$, we launch the inversion attack:
\begin{equation}
    \hat{\mX} = dec_A(\mE_V \mW).
\end{equation}

\section{Experimental Setup}

\subsection{LLMs}
We use pretrained~\textsc{FlanT5} as the backbone to launch our attack modules, encoder~$enc_{A}(\cdot)$ and decoder~$dec_{A}(\cdot)$. For victim models, a variety of encoders are experimented on, including ~\textsc{T5},~\textsc{GTR},~\textsc{mT5}, ~\textsc{mBERT} and OpenAI text embedders~\textsc{text-embedding-ada-002} (\textsc{ada-2}) and~\textsc{text-embedding-3-large} (\textsc{3-large}) (see the details of LLMs in Tabel~\ref{tab:llms}).

\subsection{Attack Experimental Setup}

\paragraph{Datasets and Attack Model}
We train the embedding-to-text generator $dec_{A}(\cdot)$ by fine-tuning ~\textsc{FlanT5}-decoder, using the MultiHPLT English dataset~\citep{de-gibert-etal-2024-new} to explore few-shot inversion attacks. 
For multilingual inversion attacks, we utilize English, German, French, and Spanish datasets from mMarco~\citep{bonifacio2021mmarco}. 
In dataset, we split 150k samples ($D_L$) to train $dec_{A}(\cdot)$; and up to 1k samples ($D_V$) to derive the alignment metrix $\mW$ by aligning $\ve_V$ to $\ve_A$, as alignment samples; and 200 for evaluation.

\begin{table}[t!]
\centering
\resizebox{\linewidth}{!}{
\begin{tabular}{c|l|c|c|c|c|c}
\toprule
 &\textbf{Model} &  \textbf{BLEU1} & \textbf{BLEU2} & \textbf{Rouge-L} & \textbf{Rouge1} & $\mathbf{COS}$ \\
\midrule

& $enc_{A}(\cdot)$ &  62.27 & 40.68 & 54.16 & 62.07 & - \\
 
\midrule
\multirow{8}{*}{\rotatebox{90}{\textbf{Vec2Text}}}  


 & \textsc{T5} (Base)  & 21.47  & 9.07 & \textbf{17.38} & 19.52  & 0.4663 \\
 & Corrector     & 18.35 & 7.60 & 15.81 & 17.76 &  \underline{0.4835}\\

 \cmidrule{2-7} 
 & \textsc{GTR}  (Base)          & 6.70 &  2.31 & 4.70 & 4.82 & 0.1911\\
 & Corrector          & 13.42 & 2.79 & \textbf{10.26} & 12.31 & \underline{0.2725}\\
 \cmidrule{2-7} 

 & \textsc{mT5} (Base)        &22.27 &  9.86 & \textbf{17.21} & 19.28 & \underline{ 0.7118}\\
 & Corrector       &18.73 &  7.79 & 15.98  & 17.82 & 0.6891 \\ 
 \cmidrule{2-7} 


& \textsc{mBERT} (Base)  &   21.56 &   9.09 &  \textbf{16.97} &18.81 & \underline{0.5335} \\
& Corrector       & 18.45 & 7.48 & 15.99 & 18.10 & 0.5531\\
\midrule

\multirow{5}{*}{\rotatebox{90}{\textbf{ALGEN}}}  

& \textsc{Random} &  11.63 & 0.6 & 7.09 & 8.36 &  -0.0440 \\
\cmidrule{2-7}

& \textsc{T5}                & 52.98 & 33.86 & \textbf{45.75} & 51.56 & \underline{0.9464} \\
& \textsc{GTR}              & 42.59 & 26.17 & 38.27 & 42.32 & 0.8879 \\
 \cmidrule{2-7} 

& \textsc{mT5}              & 49.61 & 31    & \textbf{43.35} & 48.47 & \underline{ 0.9370} \\
& \textsc{mBERT}            & 47.06 & 28.66 & 39.9  & 45.04 & 0.9217 \\
 \cmidrule{2-7} 
& \textsc{OpenAI (ada-2)} &  46.7  & 28.67 & \textbf{41.45} & 47.01 & \underline{0.9312} \\  
&  \textsc{OpenAI (3-large)}     & 46.28 & 28.74 & 41.31 & 46.28 & 0.9066 \\

\bottomrule
\end{tabular}}
\caption{Inversion Attack Performances by victim models with 1,000 leaked data samples. The best Rouge-L scores are \textbf{bolded}, and the highest cosine similarities are \underline{underlined}.}
\label{tab:inversion_attack_performance}
\end{table}

\begin{table}[h!]
\centering
\resizebox{\linewidth}{!}{
\begin{tabular}{l|ccrrr}
    \toprule
    Dataset & Clasisification & \#Class  & \#Train & \#Dev & \#Test \\
    \midrule
    SNLI & NLI & 3 & 540,000 & 200 & 200 \\
        SST2 & Sentiment & 2 & 59,560 &  200 & 200 \\
    S140 & Sentiment & 2 & 1,599,798 & 200 &200\\
    \bottomrule
    \end{tabular}
    }
    \caption{Statistics of Utility Task Datasets}
\label{tab:utility_stats}
\end{table}

\paragraph{Embedding-to-Text Generator Training}
We conduct a series of experiments varying the training set size (from 10k to 1M samples), learning rate and weight decay, then select the generator configuration that achieves the best Rouge-L score on the evaluation set.
Eventually, to strike a balance of performance and data usage, we train $dec_{A}(\cdot)$, an embedding-to-sequence decoder, with a learning rate of $1e-4$ and weight decay $1e-4$ on AdamW optimizer~\citep{loshchilov2017fixing}, batch size 128, and 150k data samples performs the best.
We use \text{Cross-entropy Loss} for training the generator.


\paragraph{Evaluation Metrics}
~\textbf{Rouge-L}~\citep{lin-2004-rouge} is used to measure the accuracy and overlap between ground truth text $x$ and reconstructed text $\hat{x}$ based on n-grams. 
In addition, we report~\textbf{Rouge1},~\textbf{BLEU1} and ~\textbf{BLEU2}.
~\textbf{Cosine Similarity (COS)} between the aligned victim embeddings $\ve_{V\rightarrow A}$ and the attack embeddings $\ve_{A}$ is calculated to evaluate the semantic similarity in the latent embedding space.

\subsection{Defense Experimental Setup}~\label{sec:experiment_utility}
We aim to evaluate how embeddings with defense mechanisms perform in downstream tasks.

\paragraph{Datasets}
We use SST2~\citep{socher-etal-2013-recursive}, sentiment140 (S140)~\citep{go2009twitter} and SNLI~\citep{bowman-etal-2015-large} in our experiments, which are curated to ensure a balanced distribution of labels.
Table~\ref{tab:utility_stats} shows the statistics of datasets.

\paragraph{Utility of Embeddings}
Using the embeddings from the victim encoders, we train multi-layer perception classifiers on the datasets and evaluate the accuracy (ACC) and \fscore (F1) performance. 
We train each classifier 6 epochs, and select the best model with ACC on the dev dataset to evaluate the test dataset.
There would be minimal difference in the utility performance between the protected and original embeddings, if a defense is successful.

\section{Few-shot Inversion Attacks}
Each subsection aims to answer one Research Question (RQ).

\subsection{How few Leaked Data do Attackers Need?}

\begin{figure}[t!]
    \centering
    \includegraphics[width=\linewidth]{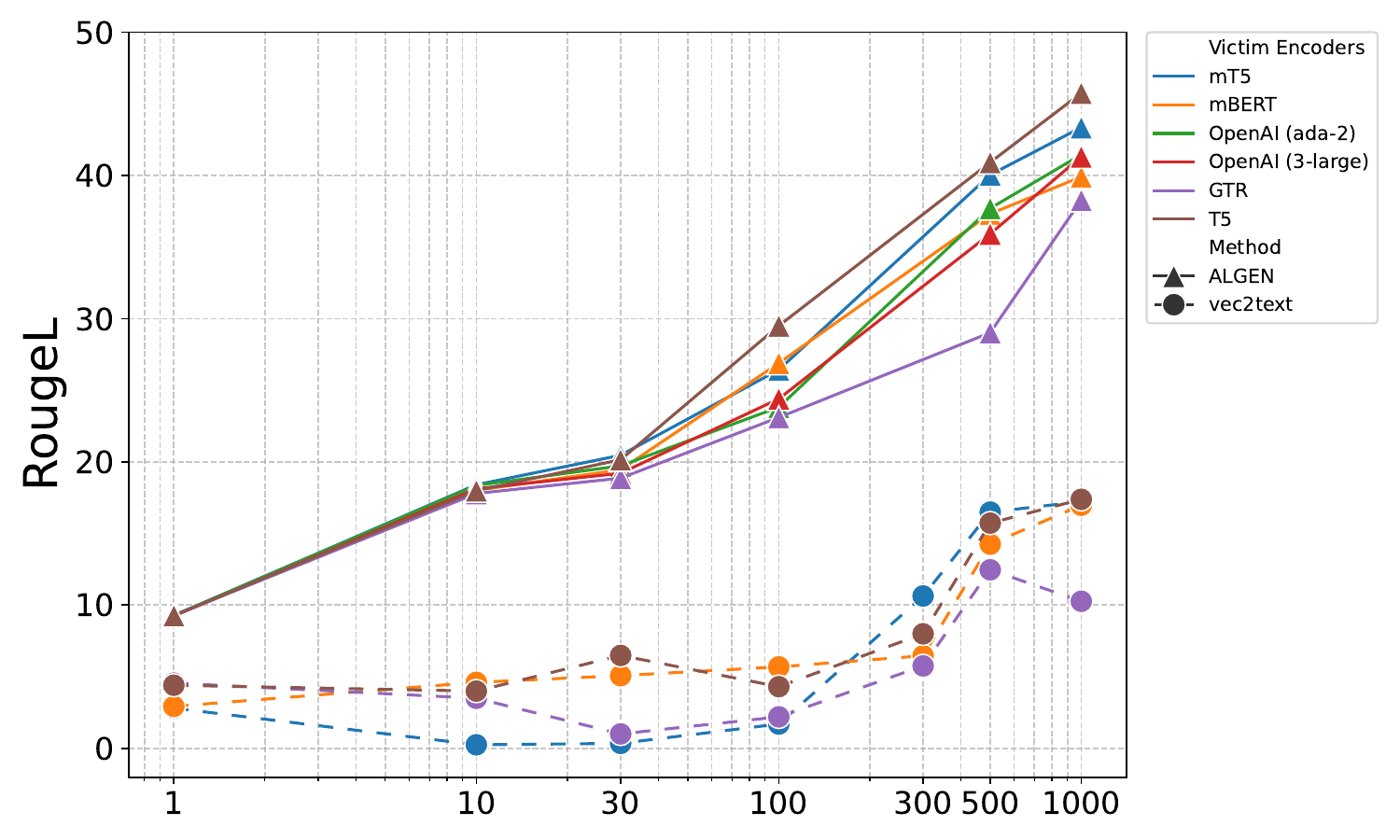}
    \includegraphics[width=\linewidth]{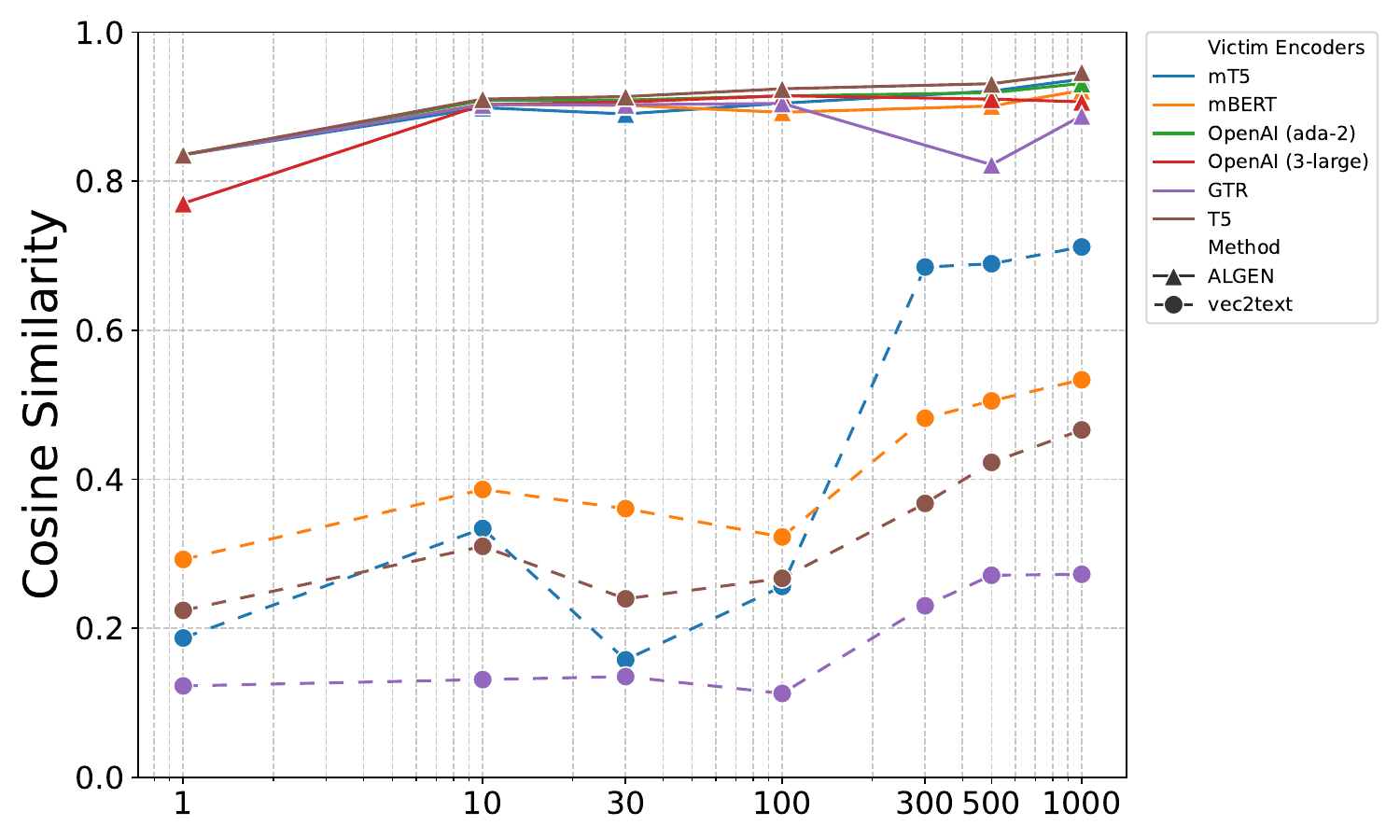}
    \caption{Inversion Performance in Rouge-L (Top) and Cosine Similarities (Bottom) by Victim Models and Alignment samples. Dashed lines are results of Vec2Text and solid lines are results of \textbf{ALGEN}.}
    \label{fig:inversion_cos_rougel}
\end{figure}

With only a single leaked data sample, our attack model manages to invert the victim embeddings, achieving a Rouge-L score of 10 across the encoders, as shown in Fig.~\ref{fig:inversion_cos_rougel}. 
We use randomly generated embeddings as a baseline to verify that the aligned embeddings from $\textbf{\ourmethod}$ capture meaningful information. 
As shown in Table~\ref{tab:inversion_attack_performance}, all victim embeddings substantially outperform the \textsc{random} across metrics, validating our approach.
Furthermore, when the number of leaked data samples increases until 1k, the inversion performance increases sharply, reaching 45.75 in Rouge-L and $0.9464$ for cosine similarity for \textsc{T5} embeddings.
Notably, while \textsc{GTR} and \textsc{T5} share the same tokenizer with the attack model, their inversion performances are not superior to others. 
Moreover, the inversion performance on proprietary \textsc{OpenAI} embeddings also reaches comparable performance, more than 41 in Rouge-L and 0.9 in cosine similarities for both~\textsc{ada-2} and~\textsc{3-large}, highlighting the risks posed by inversion attacks.
As shown in the qualitative analysis in Table~\ref{tab:qualitative}, some sentences can be inverted with almost an exact match.
Furthermore, we conduct an ablation study with the size of alignment data samples and find that 1k alignment samples strike a balance between data size and performances (see  Fig.~\ref{fig:datasize_ablation} in Appendix.~\ref{apendix:datasize}).

We compare our method with Vec2Text~\citep{morris2023text}, which trains two-step models (i.e., Base and Corrector) with iterative access to the victim encoder, and it requires training on embeddings from each encoder to invert the regarding embeddings.
In comparison, our method only requires training one local attack model, and training does not involve specific victim encoders.
Up to 1k samples, Vec2Text performance is much inferior compared to our method both in Rouge-L and cosine similarities, lower than 20 and 0.72, respectively, as detailed in Fig.~\ref{fig:inversion_cos_rougel} and Table~\ref{tab:inversion_attack_performance}.

We generally observe small alignment errors, with the cosine similarities between victim embeddings and attack embeddings consistently higher than 0.8 and near 1.0 when the number of alignment samples increases. 
The bottleneck of the performance of \textbf{ALGEN} is likely to lie in the decoding, as the trained attack decoder can only reach 54.16 in Rouge-L to invert the attack embeddings, as shown in Table~\ref{tab:inversion_attack_performance},
which is considered to be the upper bound of inversion attack performance.

\subsection{Are Other Languages (More) Vulnerable? }
Building on previous work on multilingual embedding inversion, we also investigate the impact of \textbf{\ourmethod} on languages other than English.
To achieve this, we trained local attack models in English, French, German, and Spanish. 
We further conduct crosslingual embedding inversion attacks, for example, applying the English-trained attack model to invert French textual embeddings. 
Table~\ref{tab:crosslingual_inversion} summarizes the results of multilingual and crosslingual embedding inversion attacks. 
The first row shows the results of monolingual inversion on the attack embeddings, which serve as an upper bound of the inversion performances on this dataset. 
As expected, the monolingual inversion attacks perform better than the crosslingual ones.

\begin{table}[t!]
    \centering
    \resizebox{\linewidth}{!}{

    \begin{tabular}{c|c|l|l|l|l}

\toprule
      & \multirow{2}{*}{\textbf{Attack Lang.}} & \multicolumn{4}{c}{\textbf{Victim Languages}}\\
      
       &   & \multicolumn{1}{c}{\textbf{English}} &\multicolumn{1}{c}{\textbf{French}} & \multicolumn{1}{c}{\textbf{German}} & \textbf{Spanish} \\ 
       \midrule      
    $enc_{A}(\cdot)$ &    & 54.47 & 52.78 & 24.77 & 53.98 \\
 \midrule
       
       \multirow{4}{*}{ {\textsc{T5}}} 

       &  \textbf{English} & 29.29 & 12.57 (+2.58) & 5.54 (+0.7) & 13.31 (+0.33) \\ 
       &  \textbf{French} & 8.67 (+3.37) & 31.01 & 2.4 (+1.25) & 13.72 (+0.84) \\ 
       &  \textbf{German} & 15.72 (+0.09) & 15.56 (+0.06) & 13.04 & 16.63 (+0.09) \\ 
       &  \textbf{Spanish} & 6.56 (+4.49) & 11.33 (+1.05) & 2.18 (+0.95) & 31.83 \\ 
        \midrule
        
          \multirow{4}{*}{\textsc{GTR}} 

       &  \textbf{English} & 19.22 & 10.45 (+1.34) & 4.27 (+0.57) & 10.17 (+1.09) \\ 
       &  \textbf{French} & 4.78 (+2.85) & 23.98 & 1.95 (+1.14) & 12.16 (+0.42) \\ 
       &  \textbf{German} & 11.13 (+0.48) & 13.37 (-0.15) & 8.48 & 14.32 (-0.09) \\ 
      &   \textbf{Spanish} & 4.4 (+3.74) & 10.5 (+1.98) & 1.9 (+0.95) & 20.88 \\ 
        \midrule
        
 \multirow{4}{*}{\textsc{mT5}} 
 
 &  \textbf{English} & 25.48 & 12.29 (+1.64) & 5.31 (+0.03) & 12.61 (+0.87) \\ 
  &  \textbf{French} & 8.66 (+2.79) & 24.6 & 2.27 (+1.43) & 13.1 (+0.48) \\ 
        &  \textbf{German} & 15.54 (+0.06) & 15.2 (-0.1) & 10.09 & 15.59 (+0.02) \\ 
       &  \textbf{Spanish} & 6.84 (+3.98) & 11.62 (+1.35) & 1.88 (+1.31) & 24.05 \\ 

       \midrule
 \multirow{4}{*}{\textsc{mBERT}} 
       & \textbf{English} & 21.3 & 11.72 (+1.42) & 4.46 (+0.3) & 11.76 (+0.82) \\ 
      &  \textbf{French} & 5.91 (+3.61) & 22.79 & 1.87 (+1.71) & 12.05 (-0.19) \\ 
      &  \textbf{German} & 14.29 (+0.12) & 14.16 (-0.05) & 9.17 & 15.6 (+0.05) \\ 
      &  \textbf{Spanish} & 5.17 (+4.14) & 11.07 (+0.09) & 1.75 (+0.57) & 21.74 \\ 
        \midrule
    
     \multirow{4}{*}{\shortstack{\textsc{OpenAI}\\ \textsc{(ada-2)}}} 

& \textbf{English} & 24.18 & 11.62 (+1.32) & 5 (+0.17) & 11.36 (+0.95) \\ 
      &  \textbf{French} & 6.97 (+3.73) & 22.73 & 1.59 (+1.52) & 12.73 (+0.07) \\ 
      &  \textbf{German} & 14.74 (+0.11) & 13.7 (+0.07) & 9.63 & 14.33 (+0.05) \\ 
     &   \textbf{Spanish} & 7.08 (+3.57) & 9.82 (+1.39) & 1.63 (+1.05) & 21.25 \\ 
        \bottomrule
    \end{tabular}}
    \caption{Crosslingual Embedding Inversion Performance in Rouge-L with $|D_V|=$1k. The results in the brackets are the performance gain after translation.}
    \label{tab:crosslingual_inversion}
\end{table}

Consistent with previous findings~\citep{chen2024text}, crosslingual inversion often reconstructs text in a language other than the intended, usually English (the dominant language in most multilingual LLMs) or trained languages in the attack model, hindering the performance evaluation with string-match metrics.
For example, a French text ``\textit{un composé organique qui ne contient que du carbone}'', is reconstructed into English ``\textit{a chemical compound composed of carbon}''. 
Rouge-L evaluation is more accurate when the inverted text is translated back into the target language (e.g., ``\textit{un composé chimique composé de carbone}'').
To ensure fairness, we translate the inverted texts into their target languages using deep-translator.\footnote{\url{https://github.com/nidhaloff/deep-translator}}
After translation, Rouge-L scores improve across most models, with the most notable gains observed in the French-to-English scenario, as shown in Table~\ref{tab:crosslingual_inversion}.
From an attacker's perspective, splitting words in English rather than in the victim's language is advantageous. 
Attackers can be assumed to be proficient in English, while the victim's language might be unintelligible to them. This further exacerbates the vulnerabilities of non-English languages.

\subsection{Is Risk Transferable across Domains?}
To evaluate the cross-domain transferability of $\textbf{\ourmethod}$, we attack the embeddings on the mMarco English dataset using an attack model trained on MultiHPLT English data. 
Although the cross-domain inversion performance in Rouge-L is about 25\% lower than that of in-domain attacks on mMarco (see Table~\ref{tab:crosslingual_inversion}), the results remain alarmingly high - with Rouge-L near 20 and BLEU1 near 31 across victim encoders.

\begin{table}[h!]
    \centering
    \resizebox{\linewidth}{!}{
    \begin{tabular}{l|ccccc}
    \toprule
    \textbf{Model} & \textbf{BLEU1} & \textbf{BLEU2} & \textbf{Rouge-L} & \textbf{Rouge1} & \textbf{COS}  \\ 
     \midrule
        \textsc{T5} & 31.4 & 5.72 & 21.76 & 29.6 & 0.9278 \\ 
                \textsc{GTR} & 22.04 & 2.26 & 15.27 & 20.45 & 0.8442\\ 

        \textsc{mT5} & 30.08 & 4.82 & 19.33 & 26.94 & 0.9188 \\ 
           \textsc{mBERT} & 26.68 & 3.57 & 17.16 & 23.88 & 0.9033 \\ 
        \shortstack{\textsc{OpenAI (ada-2)}} & 27.62 & 4.63 & 19.07 & 26.40 & 0.9089 \\ 
         \bottomrule
    \end{tabular}}
    \caption{Cross-Domain Inversion Attack with $|D_V|$=1k.}
    \label{tab:ood}
\end{table}



\subsection{Does Inversion Recover Key Information? }

\begin{table}[h!]
    \centering
    \resizebox{\linewidth}{!}{
    \begin{tabular}{l|c|ccccccccc}
    \toprule
   \textbf{Model}  & \textbf{Overall} & \textbf{Product} & \textbf{ORG} & \textbf{GPE} & \textbf{Date} & \textbf{Time} & \textbf{Cardinal} & \textbf{Ordinal} \\ 
     \midrule
              \textsc{T5} & 23.86 & 50 & 32.41 & 21.62 & 14.55 & 16.67 & 28.57 & 40.00 \\ 

               \textsc{ mT5} & 20.2 & 53.33 & 28.04 & 17.5 & 11.11 & 0 & 10.53 & 25.00 \\ 
  \textsc{GTR} & 21.17 & 54.55 & 32.13 & 10.67 & 11.11 & 22.22 & 13.79 & 25.00 \\ 
          \textsc{mBERT} & 19.68 & 49.12 & 28.03 & 12.35 & 15.09 & 33.33 & 6.90 & 33.33 \\ 

        \textsc{OpenAI} & 22.45 & 53.33 & 34.13 & 16.47 & 13.79 & 0 & 14.55 & 37.50 \\ 
         \bottomrule
    \end{tabular}}
    \caption{Named Entity Recognition in F1 scores for overall and top 10 entities in specific. }
    \label{tab:ner}
\end{table}

To examine whether \textbf{ALGEN} attacks real key information, we apply Named Entity Recognition\footnote{\url{https://github.com/explosion/spacy-stanza}} on input and reconstructed test data from MultiHPLT English dataset to calculate the current ratio of Named Entities in the reconstructed texts. 
Table~\ref{tab:ner} shows the results in F1 scores for overall and individual entities. 
In the qualitative analysis, as shown in Table~\ref{tab:qualitative}, the input and reconstructed texts in the inversion attacks on \textsc{OpenAI (ada-2)} embeddings are compared, with named entities highlighted. 
These attacks reveal sensitive details, such as organization, country, and numbers, 
highlighting the risks of privacy disclosure by embeddings.

\begin{table}[t!]
    \centering
    \resizebox{0.9\linewidth}{!}{
    \begin{tabular}{c|c|cc|cc}
    \toprule 
        \textbf{Victim} & \textbf{Defense} & \textbf{Rouge-L}$\downarrow$ & \textbf{COS}$\downarrow$ & \textbf{ACC}$\uparrow$ & \textbf{F1}$\uparrow$ \\ 
        \midrule 
    \textsc{Random}	 & -  &	12.42	 &0.0052	 &40.5 &	40.33\\
        \midrule
          & - & 42.89 & 0.9595 & 63 & 62.92 \\ 
          \cmidrule{2-6} 
          & WET & 39.4 & 0.9562 & 63 & 62.92 \\ 
           \cmidrule{2-6} 
          & Shuffling & 35.33 & 0.9599 & 63 & 62.92 \\ 
           \cmidrule{2-6} 
         
        \textsc{T5}  & 0.001 & 43.3 & 0.9595 & 66.5 & 66.38 \\ 
          & 0.005 & 42.88 & 0.9601 & 63 & 62.55 \\ 
         & 0.01 & 40.26 & 0.9537 & 69 & 68.57 \\ 
          & 0.05 & 22.82 & 0.8459 & 59 & 58.61 \\ 
          & 0.1 & 19.32 & 0.7918 & 48 & 47.6 \\ 
          & 0.5 & 14.29 & 0.3853 & 38 & 37.92 \\ 
          & 1 & 12.97 & 0.1543 & 30.5 & 30.54 \\ 
          \midrule
         & - & 37.39 & 0.9284 & 60.5 & 60.53 \\ 
         \cmidrule{2-6} 
          & WET & 35.58 & 0.9289 & 53.5 & 53.1 \\ 
         \cmidrule{2-6} 
          & Shuffling & 31.01 & 0.9280 & 60.5 & 60.58 \\
         \cmidrule{2-6} 
        \textsc{GTR} & 0.001 & 36.52 & 0.9279 & 60 & 60.06 \\ 
          & 0.005 & 36.48 & 0.9299 & 63 & 62.91 \\ 
          & 0.01 & 34.41 & 0.9218 & 60.5 & 60.59 \\ 
          & 0.05 & 22.31 & 0.8160 & 46.5 & 46.42 \\ 
          & 0.1 & 19.39 & 0.7701 & 50.5 & 49.98 \\ 
          & 0.5 & 14.31 & 0.2920 & 30.5 & 29.97 \\ 
          & 1 & 13.48 & 0.1659 & 30.5 & 30.57 \\ 
        \midrule
          & - & 37.98 & 0.9518 & 61 & 60.95 \\ 
         \cmidrule{2-6} 
          & WET & 34.69 & 0.9479 & 55.5 & 55.12 \\ 
         \cmidrule{2-6} 
          & Shuffling & 31.83 & 0.9515 & 60.5 & 60.44 \\ 
         \cmidrule{2-6} 
        \textsc{mT5}  & 0.001 & 37.97 & 0.9519 & 61.5 & 61.47 \\ 
          & 0.005 & 37.84 & 0.9493 & 55.5 & 55.06 \\ 
          & 0.01 & 34.5 & 0.9438 & 60 & 59.96 \\ 
          & 0.05 & 21.58 & 0.8292 & 54 & 53.52 \\ 
          & 0.1 & 17.65 & 0.7578 & 36.5 & 36.45 \\ 
          & 0.5 & 14.8 & 0.4482 & 36 & 35.17 \\ 
          & 1 & 13.31 & 0.1468 & 35.5 & 34.85 \\ 
        \midrule
          & - & 35.47 & 0.9423 & 57 & 57.05 \\ 
         \cmidrule{2-6} 
          & WET & 34.35 & 0.9428 & 53.5 & 53.5 \\ 
         \cmidrule{2-6} 
          & Shuffling & 30.29 & 0.9408 & 52 & 51.93 \\ 
         \cmidrule{2-6} 
       \textsc{mBERT}   & 0.001 & 35.82 & 0.9422 & 51.5 & 51.37 \\ 
          & 0.005 & 36.18 & 0.9409 & 51 & 50.82 \\ 
          & 0.01 & 34.18 & 0.9349 & 57 & 57.01 \\ 
          & 0.05 & 22.29 & 0.8265 & 50.5 & 50.48 \\ 
          & 0.1 & 18.65 & 0.7741 & 43 & 43.03 \\ 
          & 0.5 & 13.76 & 0.3969 & 35 & 33.96 \\ 
          & 1 & 13.11 & 0.1824 & 32.5 & 32.62 \\ 
        \midrule
        & - & 38.15 & 0.9433 & 74.5 & 74.78 \\
         \cmidrule{2-6} 
         & WET & 31.74 & 0.9405 & 64.5 & 64.51 \\ 
         \cmidrule{2-6} 
          & Shuffling & 33.91 & 0.9440 & 72 & 72.31 \\ 
         \cmidrule{2-6} 
      \shortstack{\textsc{OpenAI}}   & 0.001 & 37.94 & 0.9437 & 74.5 & 74.51 \\ 
        \textsc{(ada-2)} & 0.005 & 37.63 & 0.9445 & 69 & 69.09 \\ 
       & 0.01 & 35.59 & 0.9409 & 67 & 66.85 \\ 
          & 0.05 & 20.59 & 0.8426 & 49.5 & 49.39 \\ 
         & 0.1 & 18.72 & 0.8231 & 44 & 42.76 \\ 
         & 0.5 & 14.75 & 0.4483 & 37 & 36.87 \\ 
         & 1 & 11.94 & 0.1097 & 30.5 & 30.49 \\ 
         
        \bottomrule
    \end{tabular}}
     \caption{The Inversion and Utility Performance on Classification Tasks on SNLI dataset with WET, Shuffling, Guassian Noise Insertion. From a defender's perspective, $\uparrow$ means higher are better, $\downarrow$ means lower are better.}
    \label{tab:wet_shuffling_gaussian_snli}
\end{table}

\begin{figure*}[t!]
    \centering
        \includegraphics[width=0.8\linewidth]{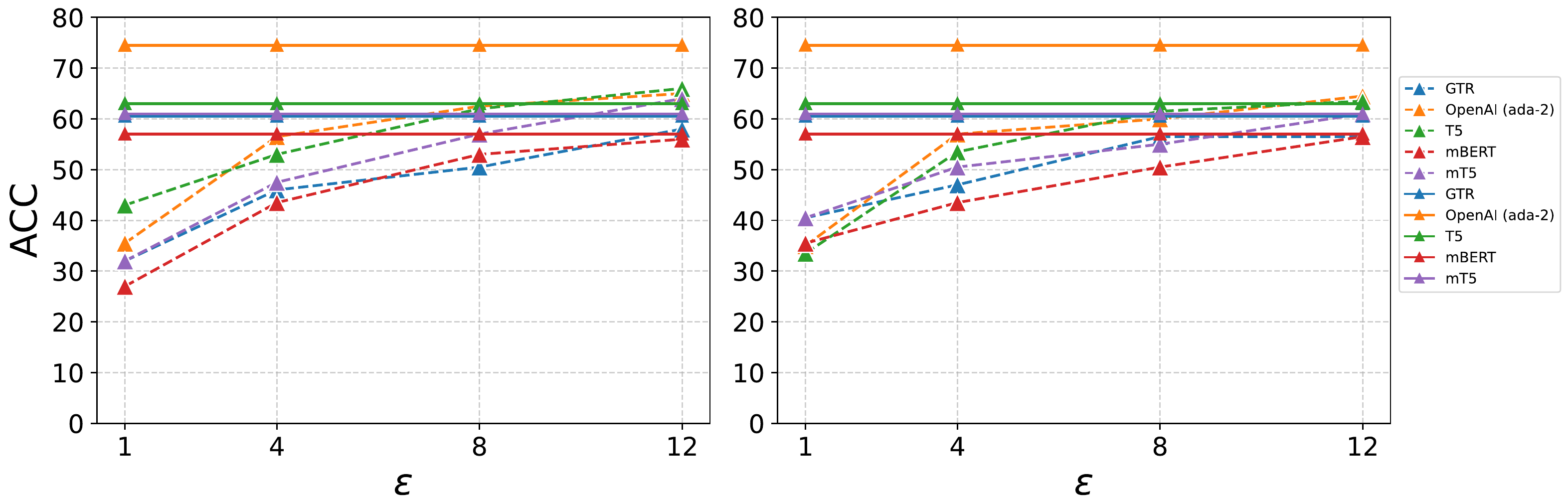} 

    \caption{The Inversion and Utility Performance in Accuracy on Classification Tasks on SNLI dataset with local DP, across $\epsilon$. The solid lines represent utility performance for non-private embeddings, while the dashed lines are for LDP-guaranteed embeddings. }
    \label{fig:ldp_snli}
\end{figure*}

\section{Defending Textual Embeddings}
To explore defenses against \textbf{\ourmethod}, we evaluate defense mechanisms designed to protect textual embeddings from adversarial attacks.

\subsection{Defense Methods}

\paragraph{WET} We implement WET on textual embeddings, to examine whether it makes embeddings robust against inversion attacks, since it is effective in defending paraphrasing attacks~\citep{shetty2024wet}. 
A transformation matrix $\mT$ is generated to transform the $\ve_{V}$ into $\ve_{WET}$ with ${\mT \cdot \ve_{V}} / {\|\mT \cdot \ve_{V}\|} $,
to ensure that i) the original elements are discarded and only the transformed ones are retained; and ii) $\mT$ is full-rank and well-conditioned to allow for accurate recovery of the original embeddings (see details of generating $\mT$ in Appendix~\ref{wet_algo}).


\paragraph{Shuffling}

We randomly shuffle the embeddings with $ \ve_{V, \pi(i)}$,
where $\pi$ is a random permutation function that reorders the indices $i$ along the hidden dimension.


\paragraph{Gaussian Noise Insertion}
We add Gaussian noises to $\ve_V$  with ${(\ve_{V}+ \lambda \cdot \epsilon)}/\|{\ve_{V}+ \lambda \cdot \epsilon}\|, \epsilon \sim \mathcal{N}(0,1)$~\citep{morris2023text,chen2024text} with $\lambda\in[0.001,0.005,0.01,0.05,0.1,0.5,1]$.

\paragraph{Differential Privacy}
~\citet{du2023sanitizing} adopts Purkayastha Mechanism (PurMech) and Normalized Planer Laplace (LapMech) on sentence embeddings to ensure metric-LDP (see details in Appendix~\ref{ldp_appendix}). 
We adopt the parameters from~\citet{du2023sanitizing} to experiment with defending textual embeddings from inversion attacks. 
The privacy budgets $\epsilon\in [1,4,8,12]$ are selected.

\begin{table}[htb!]
    \centering
     \resizebox{0.9\linewidth}{!}{
    \begin{tabular}{l|c|cc|cc}
    \toprule
   \multirow{2}{*}{\textbf{Victim}} & $\epsilon$ & \textbf{Rouge-L}$\downarrow$ & \textbf{COS}$\downarrow$ & \textbf{Rouge-L}$\downarrow$ & \textbf{COS}$\downarrow$ \\ 
\cmidrule{2-6}   
         &  & \multicolumn{2}{c|}{LapMech} & \multicolumn{2}{c}{PurMech} \\ 
        \midrule
          & 1 & 11.58 & 0.0184 & 11.38 & -0.0341 \\ 
        \textsc{T5} & 4 & 12.02 & -0.0171 & 11.52 & -0.0095 \\ 
          & 8 & 11.8 & 0.0510 & 11.96 & 0.0137 \\ 
          & 12 & 11.64 & 0.0438 & 11.48 & -0.0239 \\ 
                \midrule
          & 1 & 11.9 & 0.0327 & 11.86 & 0.0183 \\ 
        \textsc{GTR} & 4 & 11.78 & -0.0193 & 11.82 & 0.0073 \\ 
          & 8 & 11.89 & -0.0217 & 11.51 & 0.0226 \\ 
          & 12 & 10.92 & -0.0277 & 11.33 & -0.0314 \\ 
                \midrule
          & 1 & 12.3 & 0.0827 & 11.96 & 0.0143 \\ 
        \textsc{mT5} & 4 & 13.15 & 0.0696 & 13.17 & 0.0652 \\ 
          & 8 & 11.89 & -0.0148 & 12.09 & 0.0493 \\ 
          & 12 & 13.44 & 0.1179 & 11.61 & -0.0443 \\ 
                \midrule
          & 1 & 12.37 & -0.0026 & 11.33 & 0.0016 \\ 
        \textsc{mBERT} & 4 & 11.84 & -0.0431 & 11.58 & -0.0019 \\ 
          & 8 & 11.58 & -0.0333 & 12.93 & 0.1003 \\ 
          & 12 & 11 & -0.0242 & 11.48 & 0.0271 \\ 
        \midrule
        & 1 & 12.33 & 0.0198 & 11.15 & -0.0020 \\ 
       \shortstack{\textsc{OpenAI}}& 4 & 12.31 & 0.0133 & 11.87 & -0.0417 \\ 
        \textsc{(ada-2)} & 8 & 10.87 & -0.0659 & 13.13 & 0.1078 \\ 
        & 12 & 11.07 & -0.0050 &12.48 & 0.0447 \\ 
        \bottomrule
    \end{tabular}
    }
     \caption{The Inversion Performance on Classification Tasks on SNLI dataset with Local DP. From a defender's perspective, $\downarrow$ means lower are better.}
         \label{tab:snli_ldp_inversion}

\end{table}

\subsection{Results}
As shown in Table~\ref{tab:wet_shuffling_gaussian_snli},~\ref{tab:wet_shuffling_gaussian_sst2} and~\ref{tab:wet_shuffling_gaussian_s140}, WET and Shuffling have minimal impact on both inversion attack performance and the utility performance across victim models and datasets.
The randomly generated embeddings are also used as a baseline.
With Gaussian noise insertion, the bigger the noise $\lambda$, both performance in inversion and utility decrease.
Using local DP, while the utility performance is preserved almost as the non-private embeddings with $\epsilon=12$, as shown in Fig.~\ref{fig:ldp_snli}, Tabel~\ref{tab:ldp_sst2} and~\ref{tab:s140_ldp}. 
However, the inversion performance still maintains more than 25\% of the non-private embeddings in Rouge-L across encoders for both LapMech and PurMech, as detailed in Table~\ref{tab:snli_ldp_inversion},~\ref{tab:ldp_sst2} and~\ref{tab:s140_ldp}, posing security and privacy risks for the embeddings.

\section{Discussion and Conclusion}
In this work, we introduce and validate the effectiveness of a novel few-shot inversion attack,~\textbf{\ourmethod}, 
which drastically lowers the cost and complexity of such attacks on widely used vector databases.
Our results show that the attack transfers effectively across domains and languages while revealing critical information.
Moreover, its ability to align embeddings from different LLMs with minimal loss highlights its broad NLP applications, especially in cross-lingual embedding alignment.
Finally, our evaluation of existing defense mechanisms reveals that none can adequately protect textual embeddings from inversion attacks while maintaining utility, highlighting significant security and privacy vulnerabilities.







\section*{Limitations}
Our work does not propose a sufficient defense mechanism for \textbf{\ourmethod}. 
Although we evaluated a number of existing defense mechanisms for textual embeddings, we found them to be ineffective against the proposed embedding inversion attack. The primary focus of this work is to expose the security vulnerabilities in embedding services and to inspire the development of future defense paradigms.

\section*{Computational Resources}
We conduct experiments and train each text-to-embedding generator model on a single Nvidia A40 GPU, with the training process completing in three hours. 
Beyond this, \textbf{\ourmethod} requires minimal GPU resources, making it a genuinely few-shot experimental setting.

\section*{Ethics Statement}
We comply with the ACL Ethics Policy. 
The inversion attacks implemented in this paper can be misused and potentially harmful to proprietary embeddings.
We discuss and experiment with potential mitigation and defense mechanisms, and we encourage further research in developing effective defenses in this attack space.

\section*{Acknowledgements}
YC and JB are funded by the Carlsberg Foundation, under the Semper Ardens: Accelerate programme (project nr. CF21-0454). We further acknowledge the support of the AAU AI Cloud and express our gratitude to DeiC for providing computing resources on the LUMI cluster (project nr. DeiC-AAU-N5-2024085-H2-2024-28).
YC expresses her gratitude to the Danish Ministry of Higher Education and Science for the EliteForsk travel grant and to the School of Computing at Macquarie University for hosting her research stay.
QK acknowledges support from 24 FSE DDRI Grant and 2024 FSE Strategic Startup.




\bibliography{anthology,custom}

\begin{thebibliography}{48}
\expandafter\ifx\csname natexlab\endcsname\relax\def\natexlab#1{#1}\fi

\bibitem[{Aboagye et~al.(2022)Aboagye, Zheng, Yeh, Wang, Zhang, Wang, Yang, and Phillips}]{aboagye2022normalization}
Prince~Osei Aboagye, Yan Zheng, Chin-Chia~Michael Yeh, Junpeng Wang, Wei Zhang, Liang Wang, Hao Yang, and Jeff~M Phillips. 2022.
\newblock Normalization of language embeddings for cross-lingual alignment.
\newblock In \emph{International Conference on Learning Representations}.

\bibitem[{Aldarmaki and Diab(2019)}]{aldarmaki-diab-2019-context}
Hanan Aldarmaki and Mona Diab. 2019.
\newblock \href {https://doi.org/10.18653/v1/N19-1391} {Context-aware cross-lingual mapping}.
\newblock In \emph{Proceedings of the 2019 Conference of the North {A}merican Chapter of the Association for Computational Linguistics: Human Language Technologies, Volume 1 (Long and Short Papers)}, pages 3906--3911, Minneapolis, Minnesota. Association for Computational Linguistics.

\bibitem[{Alqahtani et~al.(2021)Alqahtani, Lalwani, Zhang, Romeo, and Mansour}]{alqahtani-etal-2021-using-optimal}
Sawsan Alqahtani, Garima Lalwani, Yi~Zhang, Salvatore Romeo, and Saab Mansour. 2021.
\newblock \href {https://doi.org/10.18653/v1/2021.findings-emnlp.329} {Using optimal transport as alignment objective for fine-tuning multilingual contextualized embeddings}.
\newblock In \emph{Findings of the Association for Computational Linguistics: EMNLP 2021}, pages 3904--3919, Punta Cana, Dominican Republic. Association for Computational Linguistics.

\bibitem[{Artetxe et~al.(2016)Artetxe, Labaka, and Agirre}]{artetxe-etal-2016-learning}
Mikel Artetxe, Gorka Labaka, and Eneko Agirre. 2016.
\newblock \href {https://doi.org/10.18653/v1/D16-1250} {Learning principled bilingual mappings of word embeddings while preserving monolingual invariance}.
\newblock In \emph{Proceedings of the 2016 Conference on Empirical Methods in Natural Language Processing}, pages 2289--2294, Austin, Texas. Association for Computational Linguistics.

\bibitem[{Artetxe et~al.(2017)Artetxe, Labaka, and Agirre}]{artetxe-etal-2017-learning}
Mikel Artetxe, Gorka Labaka, and Eneko Agirre. 2017.
\newblock \href {https://doi.org/10.18653/v1/P17-1042} {Learning bilingual word embeddings with (almost) no bilingual data}.
\newblock In \emph{Proceedings of the 55th Annual Meeting of the Association for Computational Linguistics (Volume 1: Long Papers)}, pages 451--462, Vancouver, Canada. Association for Computational Linguistics.

\bibitem[{Bonifacio et~al.(2021)Bonifacio, Jeronymo, Abonizio, Campiotti, Fadaee, Lotufo, and Nogueira}]{bonifacio2021mmarco}
Luiz Bonifacio, Vitor Jeronymo, Hugo~Queiroz Abonizio, Israel Campiotti, Marzieh Fadaee, Roberto Lotufo, and Rodrigo Nogueira. 2021.
\newblock mmarco: A multilingual version of the ms marco passage ranking dataset.
\newblock \emph{arXiv preprint arXiv:2108.13897}.

\bibitem[{Bowman et~al.(2015)Bowman, Angeli, Potts, and Manning}]{bowman-etal-2015-large}
Samuel~R. Bowman, Gabor Angeli, Christopher Potts, and Christopher~D. Manning. 2015.
\newblock \href {https://doi.org/10.18653/v1/D15-1075} {A large annotated corpus for learning natural language inference}.
\newblock In \emph{Proceedings of the 2015 Conference on Empirical Methods in Natural Language Processing}, pages 632--642, Lisbon, Portugal. Association for Computational Linguistics.

\bibitem[{Brown et~al.(2020)Brown, Mann, Ryder, Subbiah, Kaplan, Dhariwal, Neelakantan, Shyam, Sastry, Askell, Agarwal, Herbert-Voss, Krueger, Henighan, Child, Ramesh, Ziegler, Wu, Winter, Hesse, Chen, Sigler, Litwin, Gray, Chess, Clark, Berner, McCandlish, Radford, Sutskever, and Amodei}]{brown2020languagemodelsfewshotlearners}
Tom~B. Brown, Benjamin Mann, Nick Ryder, Melanie Subbiah, Jared Kaplan, Prafulla Dhariwal, Arvind Neelakantan, Pranav Shyam, Girish Sastry, Amanda Askell, Sandhini Agarwal, Ariel Herbert-Voss, Gretchen Krueger, Tom Henighan, Rewon Child, Aditya Ramesh, Daniel~M. Ziegler, Jeffrey Wu, Clemens Winter, Christopher Hesse, Mark Chen, Eric Sigler, Mateusz Litwin, Scott Gray, Benjamin Chess, Jack Clark, Christopher Berner, Sam McCandlish, Alec Radford, Ilya Sutskever, and Dario Amodei. 2020.
\newblock \href {http://arxiv.org/abs/2005.14165} {Language models are few-shot learners}.

\bibitem[{Cao et~al.(2020)Cao, Kitaev, and Klein}]{cao2020multilingualalignmentcontextualword}
Steven Cao, Nikita Kitaev, and Dan Klein. 2020.
\newblock \href {http://arxiv.org/abs/2002.03518} {Multilingual alignment of contextual word representations}.

\bibitem[{Chen et~al.(2024{\natexlab{a}})Chen, Biswas, Lent, and Bjerva}]{chen2024typ}
Yiyi Chen, Russa Biswas, Heather Lent, and Johannes Bjerva. 2024{\natexlab{a}}.
\newblock Against all odds: Overcoming typology, script, and language confusion in multilingual embedding inversion attacks.
\newblock \emph{arXiv preprint arXiv:2408.11749}.

\bibitem[{Chen et~al.(2024{\natexlab{b}})Chen, Lent, and Bjerva}]{chen2024text}
Yiyi Chen, Heather Lent, and Johannes Bjerva. 2024{\natexlab{b}}.
\newblock Text embedding inversion security for multilingual language models.
\newblock In \emph{Proceedings of the 62nd Annual Meeting of the Association for Computational Linguistics (Volume 1: Long Papers)}, pages 7808--7827.

\bibitem[{Chung et~al.(2022)Chung, Hou, Longpre, Zoph, Tay, Fedus, Li, Wang, Dehghani, Brahma, Webson, Gu, Dai, Suzgun, Chen, Chowdhery, Castro-Ros, Pellat, Robinson, Valter, Narang, Mishra, Yu, Zhao, Huang, Dai, Yu, Petrov, Chi, Dean, Devlin, Roberts, Zhou, Le, and Wei}]{chung2022scalinginstructionfinetunedlanguagemodels}
Hyung~Won Chung, Le~Hou, Shayne Longpre, Barret Zoph, Yi~Tay, William Fedus, Yunxuan Li, Xuezhi Wang, Mostafa Dehghani, Siddhartha Brahma, Albert Webson, Shixiang~Shane Gu, Zhuyun Dai, Mirac Suzgun, Xinyun Chen, Aakanksha Chowdhery, Alex Castro-Ros, Marie Pellat, Kevin Robinson, Dasha Valter, Sharan Narang, Gaurav Mishra, Adams Yu, Vincent Zhao, Yanping Huang, Andrew Dai, Hongkun Yu, Slav Petrov, Ed~H. Chi, Jeff Dean, Jacob Devlin, Adam Roberts, Denny Zhou, Quoc~V. Le, and Jason Wei. 2022.
\newblock \href {http://arxiv.org/abs/2210.11416} {Scaling instruction-finetuned language models}.

\bibitem[{Conneau et~al.(2018)Conneau, Lample, Rinott, Williams, Bowman, Schwenk, and Stoyanov}]{conneau2018xnlievaluatingcrosslingualsentence}
Alexis Conneau, Guillaume Lample, Ruty Rinott, Adina Williams, Samuel~R. Bowman, Holger Schwenk, and Veselin Stoyanov. 2018.
\newblock \href {http://arxiv.org/abs/1809.05053} {Xnli: Evaluating cross-lingual sentence representations}.

\bibitem[{de~Gibert et~al.(2024)de~Gibert, Nail, Arefyev, Ba{\~n}{\'o}n, van~der Linde, Ji, Zaragoza-Bernabeu, Aulamo, Ram{\'i}rez-S{\'a}nchez, Kutuzov, Pyysalo, Oepen, and Tiedemann}]{de-gibert-etal-2024-new}
Ona de~Gibert, Graeme Nail, Nikolay Arefyev, Marta Ba{\~n}{\'o}n, Jelmer van~der Linde, Shaoxiong Ji, Jaume Zaragoza-Bernabeu, Mikko Aulamo, Gema Ram{\'i}rez-S{\'a}nchez, Andrey Kutuzov, Sampo Pyysalo, Stephan Oepen, and J{\"o}rg Tiedemann. 2024.
\newblock \href {https://aclanthology.org/2024.lrec-main.100/} {A new massive multilingual dataset for high-performance language technologies}.
\newblock In \emph{Proceedings of the 2024 Joint International Conference on Computational Linguistics, Language Resources and Evaluation (LREC-COLING 2024)}, pages 1116--1128, Torino, Italia. ELRA and ICCL.

\bibitem[{Devlin(2018)}]{devlin2018bert}
Jacob Devlin. 2018.
\newblock Bert: Pre-training of deep bidirectional transformers for language understanding.
\newblock \emph{arXiv preprint arXiv:1810.04805}.

\bibitem[{Devlin et~al.(2019)Devlin, Chang, Lee, and Toutanova}]{devlin2019bertpretrainingdeepbidirectional}
Jacob Devlin, Ming-Wei Chang, Kenton Lee, and Kristina Toutanova. 2019.
\newblock \href {http://arxiv.org/abs/1810.04805} {Bert: Pre-training of deep bidirectional transformers for language understanding}.

\bibitem[{Du et~al.(2023)Du, Yue, Chow, and Sun}]{du2023sanitizing}
Minxin Du, Xiang Yue, Sherman~SM Chow, and Huan Sun. 2023.
\newblock Sanitizing sentence embeddings (and labels) for local differential privacy.
\newblock In \emph{Proceedings of the ACM Web Conference 2023}, pages 2349--2359.

\bibitem[{Dwork et~al.(2014)Dwork, Roth et~al.}]{dwork2014algorithmic}
Cynthia Dwork, Aaron Roth, et~al. 2014.
\newblock The algorithmic foundations of differential privacy.
\newblock \emph{Foundations and Trends{\textregistered} in Theoretical Computer Science}, 9(3--4):211--407.

\bibitem[{Go et~al.(2009)Go, Bhayani, and Huang}]{go2009twitter}
Alec Go, Richa Bhayani, and Lei Huang. 2009.
\newblock Twitter sentiment classification using distant supervision.
\newblock \emph{CS224N project report, Stanford}, 1(12):2009.

\bibitem[{Gray et~al.(2006)}]{gray2006toeplitz}
Robert~M Gray et~al. 2006.
\newblock Toeplitz and circulant matrices: A review.
\newblock \emph{Foundations and Trends{\textregistered} in Communications and Information Theory}, 2(3):155--239.

\bibitem[{Huang et~al.(2024)Huang, Tsai, Hsiao, Lin, and Lin}]{huang_transferable_2024}
Yu-Hsiang Huang, Yuche Tsai, Hsiang Hsiao, Hong-Yi Lin, and Shou-De Lin. 2024.
\newblock \href {https://doi.org/10.18653/v1/2024.acl-long.230} {Transferable {Embedding} {Inversion} {Attack}: {Uncovering} {Privacy} {Risks} in {Text} {Embeddings} without {Model} {Queries}}.
\newblock In \emph{Proceedings of the 62nd {Annual} {Meeting} of the {Association} for {Computational} {Linguistics} ({Volume} 1: {Long} {Papers})}, pages 4193--4205, Bangkok, Thailand. Association for Computational Linguistics.

\bibitem[{Jalili~Sabet et~al.(2020)Jalili~Sabet, Dufter, Yvon, and Sch{\"u}tze}]{jalili-sabet-etal-2020-simalign}
Masoud Jalili~Sabet, Philipp Dufter, Fran{\c{c}}ois Yvon, and Hinrich Sch{\"u}tze. 2020.
\newblock \href {https://doi.org/10.18653/v1/2020.findings-emnlp.147} {{S}im{A}lign: High quality word alignments without parallel training data using static and contextualized embeddings}.
\newblock In \emph{Findings of the Association for Computational Linguistics: EMNLP 2020}, pages 1627--1643, Online. Association for Computational Linguistics.

\bibitem[{Kasiviswanathan et~al.(2011)Kasiviswanathan, Lee, Nissim, Raskhodnikova, and Smith}]{kasiviswanathan2011can}
Shiva~Prasad Kasiviswanathan, Homin~K Lee, Kobbi Nissim, Sofya Raskhodnikova, and Adam Smith. 2011.
\newblock What can we learn privately?
\newblock \emph{SIAM Journal on Computing}, 40(3):793--826.

\bibitem[{Krahn et~al.(2023)Krahn, Tate, and Lamicela}]{krahn2023sentenceembeddingmodelsancient}
Kevin Krahn, Derrick Tate, and Andrew~C. Lamicela. 2023.
\newblock \href {http://arxiv.org/abs/2308.13116} {Sentence embedding models for ancient greek using multilingual knowledge distillation}.

\bibitem[{Lample et~al.(2018)Lample, Conneau, Ranzato, Denoyer, and J{\'e}gou}]{lample2018word}
Guillaume Lample, Alexis Conneau, Marc'Aurelio Ranzato, Ludovic Denoyer, and Herv{\'e} J{\'e}gou. 2018.
\newblock Word translation without parallel data.
\newblock In \emph{International conference on learning representations}.

\bibitem[{Lewis et~al.(2020)Lewis, Perez, Piktus, Petroni, Karpukhin, Goyal, K\"{u}ttler, Lewis, Yih, Rockt\"{a}schel, Riedel, and Kiela}]{rag_lewis_2020}
Patrick Lewis, Ethan Perez, Aleksandra Piktus, Fabio Petroni, Vladimir Karpukhin, Naman Goyal, Heinrich K\"{u}ttler, Mike Lewis, Wen-tau Yih, Tim Rockt\"{a}schel, Sebastian Riedel, and Douwe Kiela. 2020.
\newblock Retrieval-augmented generation for knowledge-intensive nlp tasks.
\newblock In \emph{Proceedings of the 34th International Conference on Neural Information Processing Systems}, NIPS '20, Red Hook, NY, USA. Curran Associates Inc.

\bibitem[{Li et~al.(2023)Li, Xu, and Song}]{li-etal-2023-sentence}
Haoran Li, Mingshi Xu, and Yangqiu Song. 2023.
\newblock \href {https://doi.org/10.18653/v1/2023.findings-acl.881} {Sentence embedding leaks more information than you expect: Generative embedding inversion attack to recover the whole sentence}.
\newblock In \emph{Findings of the Association for Computational Linguistics: ACL 2023}, pages 14022--14040, Toronto, Canada. Association for Computational Linguistics.

\bibitem[{Lin(2004)}]{lin-2004-rouge}
Chin-Yew Lin. 2004.
\newblock \href {https://aclanthology.org/W04-1013} {{ROUGE}: A package for automatic evaluation of summaries}.
\newblock In \emph{Text Summarization Branches Out}, pages 74--81, Barcelona, Spain. Association for Computational Linguistics.

\bibitem[{Loshchilov et~al.(2017)Loshchilov, Hutter et~al.}]{loshchilov2017fixing}
Ilya Loshchilov, Frank Hutter, et~al. 2017.
\newblock Fixing weight decay regularization in adam.
\newblock \emph{arXiv preprint arXiv:1711.05101}, 5.

\bibitem[{Lyu et~al.(2020)Lyu, He, and Li}]{lyu-etal-2020-differentially}
Lingjuan Lyu, Xuanli He, and Yitong Li. 2020.
\newblock \href {https://doi.org/10.18653/v1/2020.findings-emnlp.213} {Differentially private representation for {NLP}: Formal guarantee and an empirical study on privacy and fairness}.
\newblock In \emph{Findings of the Association for Computational Linguistics: EMNLP 2020}, pages 2355--2365, Online. Association for Computational Linguistics.

\bibitem[{Mikolov et~al.(2013{\natexlab{a}})Mikolov, Le, and Sutskever}]{mikolov2013exploiting}
Tomas Mikolov, Quoc~V Le, and Ilya Sutskever. 2013{\natexlab{a}}.
\newblock Exploiting similarities among languages for machine translation.
\newblock \emph{arXiv preprint arXiv:1309.4168}.

\bibitem[{Mikolov et~al.(2013{\natexlab{b}})Mikolov, Sutskever, Chen, Corrado, and Dean}]{10.5555/2999792.2999959}
Tomas Mikolov, Ilya Sutskever, Kai Chen, Greg Corrado, and Jeffrey Dean. 2013{\natexlab{b}}.
\newblock Distributed representations of words and phrases and their compositionality.
\newblock In \emph{Proceedings of the 27th International Conference on Neural Information Processing Systems - Volume 2}, NIPS'13, page 3111–3119, Red Hook, NY, USA. Curran Associates Inc.

\bibitem[{Morris et~al.(2023)Morris, Kuleshov, Shmatikov, and Rush}]{morris2023text}
John~X Morris, Volodymyr Kuleshov, Vitaly Shmatikov, and Alexander~M Rush. 2023.
\newblock Text embeddings reveal (almost) as much as text.
\newblock \emph{arXiv preprint arXiv:2310.06816}.

\bibitem[{Ni et~al.(2021)Ni, Qu, Lu, Dai, Ábrego, Ma, Zhao, Luan, Hall, Chang, and Yang}]{ni2021largedualencodersgeneralizable}
Jianmo Ni, Chen Qu, Jing Lu, Zhuyun Dai, Gustavo~Hernández Ábrego, Ji~Ma, Vincent~Y. Zhao, Yi~Luan, Keith~B. Hall, Ming-Wei Chang, and Yinfei Yang. 2021.
\newblock \href {http://arxiv.org/abs/2112.07899} {Large dual encoders are generalizable retrievers}.

\bibitem[{OpenAI et~al.(2024)OpenAI, Achiam, Adler, Agarwal, Ahmad, Akkaya, Aleman, Almeida, Altenschmidt, Altman, Anadkat, Avila, Babuschkin, Balaji, Balcom, Baltescu, Bao, Bavarian, Belgum, Bello, Berdine, Bernadett-Shapiro, Berner, Bogdonoff, Boiko, Boyd, Brakman, Brockman, Brooks, Brundage, Button, Cai, Campbell, Cann, Carey, Carlson, Carmichael, Chan, Chang, Chantzis, Chen, Chen, Chen, Chen, Chen, Chess, Cho, Chu, Chung, Cummings, Currier, Dai, Decareaux, Degry, Deutsch, Deville, Dhar, Dohan, Dowling, Dunning, Ecoffet, Eleti, Eloundou, Farhi, Fedus, Felix, Fishman, Forte, Fulford, Gao, Georges, Gibson, Goel, Gogineni, Goh, Gontijo-Lopes, Gordon, Grafstein, Gray, Greene, Gross, Gu, Guo, Hallacy, Han, Harris, He, Heaton, Heidecke, Hesse, Hickey, Hickey, Hoeschele, Houghton, Hsu, Hu, Hu, Huizinga, Jain, Jain, Jang, Jiang, Jiang, Jin, Jin, Jomoto, Jonn, Jun, Kaftan, Łukasz Kaiser, Kamali, Kanitscheider, Keskar, Khan, Kilpatrick, Kim, Kim, Kim, Kirchner, Kiros, Knight, Kokotajlo, Łukasz Kondraciuk,
  Kondrich, Konstantinidis, Kosic, Krueger, Kuo, Lampe, Lan, Lee, Leike, Leung, Levy, Li, Lim, Lin, Lin, Litwin, Lopez, Lowe, Lue, Makanju, Malfacini, Manning, Markov, Markovski, Martin, Mayer, Mayne, McGrew, McKinney, McLeavey, McMillan, McNeil, Medina, Mehta, Menick, Metz, Mishchenko, Mishkin, Monaco, Morikawa, Mossing, Mu, Murati, Murk, Mély, Nair, Nakano, Nayak, Neelakantan, Ngo, Noh, Ouyang, O'Keefe, Pachocki, Paino, Palermo, Pantuliano, Parascandolo, Parish, Parparita, Passos, Pavlov, Peng, Perelman, de~Avila Belbute~Peres, Petrov, de~Oliveira~Pinto, Michael, Pokorny, Pokrass, Pong, Powell, Power, Power, Proehl, Puri, Radford, Rae, Ramesh, Raymond, Real, Rimbach, Ross, Rotsted, Roussez, Ryder, Saltarelli, Sanders, Santurkar, Sastry, Schmidt, Schnurr, Schulman, Selsam, Sheppard, Sherbakov, Shieh, Shoker, Shyam, Sidor, Sigler, Simens, Sitkin, Slama, Sohl, Sokolowsky, Song, Staudacher, Such, Summers, Sutskever, Tang, Tezak, Thompson, Tillet, Tootoonchian, Tseng, Tuggle, Turley, Tworek, Uribe, Vallone,
  Vijayvergiya, Voss, Wainwright, Wang, Wang, Wang, Ward, Wei, Weinmann, Welihinda, Welinder, Weng, Weng, Wiethoff, Willner, Winter, Wolrich, Wong, Workman, Wu, Wu, Wu, Xiao, Xu, Yoo, Yu, Yuan, Zaremba, Zellers, Zhang, Zhang, Zhao, Zheng, Zhuang, Zhuk, and Zoph}]{openai2024gpt4technicalreport}
OpenAI, Josh Achiam, Steven Adler, Sandhini Agarwal, Lama Ahmad, Ilge Akkaya, Florencia~Leoni Aleman, Diogo Almeida, Janko Altenschmidt, Sam Altman, Shyamal Anadkat, Red Avila, Igor Babuschkin, Suchir Balaji, Valerie Balcom, Paul Baltescu, Haiming Bao, Mohammad Bavarian, Jeff Belgum, Irwan Bello, Jake Berdine, Gabriel Bernadett-Shapiro, Christopher Berner, Lenny Bogdonoff, Oleg Boiko, Madelaine Boyd, Anna-Luisa Brakman, Greg Brockman, Tim Brooks, Miles Brundage, Kevin Button, Trevor Cai, Rosie Campbell, Andrew Cann, Brittany Carey, Chelsea Carlson, Rory Carmichael, Brooke Chan, Che Chang, Fotis Chantzis, Derek Chen, Sully Chen, Ruby Chen, Jason Chen, Mark Chen, Ben Chess, Chester Cho, Casey Chu, Hyung~Won Chung, Dave Cummings, Jeremiah Currier, Yunxing Dai, Cory Decareaux, Thomas Degry, Noah Deutsch, Damien Deville, Arka Dhar, David Dohan, Steve Dowling, Sheila Dunning, Adrien Ecoffet, Atty Eleti, Tyna Eloundou, David Farhi, Liam Fedus, Niko Felix, Simón~Posada Fishman, Juston Forte, Isabella Fulford, Leo
  Gao, Elie Georges, Christian Gibson, Vik Goel, Tarun Gogineni, Gabriel Goh, Rapha Gontijo-Lopes, Jonathan Gordon, Morgan Grafstein, Scott Gray, Ryan Greene, Joshua Gross, Shixiang~Shane Gu, Yufei Guo, Chris Hallacy, Jesse Han, Jeff Harris, Yuchen He, Mike Heaton, Johannes Heidecke, Chris Hesse, Alan Hickey, Wade Hickey, Peter Hoeschele, Brandon Houghton, Kenny Hsu, Shengli Hu, Xin Hu, Joost Huizinga, Shantanu Jain, Shawn Jain, Joanne Jang, Angela Jiang, Roger Jiang, Haozhun Jin, Denny Jin, Shino Jomoto, Billie Jonn, Heewoo Jun, Tomer Kaftan, Łukasz Kaiser, Ali Kamali, Ingmar Kanitscheider, Nitish~Shirish Keskar, Tabarak Khan, Logan Kilpatrick, Jong~Wook Kim, Christina Kim, Yongjik Kim, Jan~Hendrik Kirchner, Jamie Kiros, Matt Knight, Daniel Kokotajlo, Łukasz Kondraciuk, Andrew Kondrich, Aris Konstantinidis, Kyle Kosic, Gretchen Krueger, Vishal Kuo, Michael Lampe, Ikai Lan, Teddy Lee, Jan Leike, Jade Leung, Daniel Levy, Chak~Ming Li, Rachel Lim, Molly Lin, Stephanie Lin, Mateusz Litwin, Theresa Lopez, Ryan
  Lowe, Patricia Lue, Anna Makanju, Kim Malfacini, Sam Manning, Todor Markov, Yaniv Markovski, Bianca Martin, Katie Mayer, Andrew Mayne, Bob McGrew, Scott~Mayer McKinney, Christine McLeavey, Paul McMillan, Jake McNeil, David Medina, Aalok Mehta, Jacob Menick, Luke Metz, Andrey Mishchenko, Pamela Mishkin, Vinnie Monaco, Evan Morikawa, Daniel Mossing, Tong Mu, Mira Murati, Oleg Murk, David Mély, Ashvin Nair, Reiichiro Nakano, Rajeev Nayak, Arvind Neelakantan, Richard Ngo, Hyeonwoo Noh, Long Ouyang, Cullen O'Keefe, Jakub Pachocki, Alex Paino, Joe Palermo, Ashley Pantuliano, Giambattista Parascandolo, Joel Parish, Emy Parparita, Alex Passos, Mikhail Pavlov, Andrew Peng, Adam Perelman, Filipe de~Avila Belbute~Peres, Michael Petrov, Henrique~Ponde de~Oliveira~Pinto, Michael, Pokorny, Michelle Pokrass, Vitchyr~H. Pong, Tolly Powell, Alethea Power, Boris Power, Elizabeth Proehl, Raul Puri, Alec Radford, Jack Rae, Aditya Ramesh, Cameron Raymond, Francis Real, Kendra Rimbach, Carl Ross, Bob Rotsted, Henri Roussez,
  Nick Ryder, Mario Saltarelli, Ted Sanders, Shibani Santurkar, Girish Sastry, Heather Schmidt, David Schnurr, John Schulman, Daniel Selsam, Kyla Sheppard, Toki Sherbakov, Jessica Shieh, Sarah Shoker, Pranav Shyam, Szymon Sidor, Eric Sigler, Maddie Simens, Jordan Sitkin, Katarina Slama, Ian Sohl, Benjamin Sokolowsky, Yang Song, Natalie Staudacher, Felipe~Petroski Such, Natalie Summers, Ilya Sutskever, Jie Tang, Nikolas Tezak, Madeleine~B. Thompson, Phil Tillet, Amin Tootoonchian, Elizabeth Tseng, Preston Tuggle, Nick Turley, Jerry Tworek, Juan Felipe~Cerón Uribe, Andrea Vallone, Arun Vijayvergiya, Chelsea Voss, Carroll Wainwright, Justin~Jay Wang, Alvin Wang, Ben Wang, Jonathan Ward, Jason Wei, CJ~Weinmann, Akila Welihinda, Peter Welinder, Jiayi Weng, Lilian Weng, Matt Wiethoff, Dave Willner, Clemens Winter, Samuel Wolrich, Hannah Wong, Lauren Workman, Sherwin Wu, Jeff Wu, Michael Wu, Kai Xiao, Tao Xu, Sarah Yoo, Kevin Yu, Qiming Yuan, Wojciech Zaremba, Rowan Zellers, Chong Zhang, Marvin Zhang, Shengjia
  Zhao, Tianhao Zheng, Juntang Zhuang, William Zhuk, and Barret Zoph. 2024.
\newblock \href {http://arxiv.org/abs/2303.08774} {Gpt-4 technical report}.

\bibitem[{Radford(2018)}]{radford2018improving}
Alec Radford. 2018.
\newblock Improving language understanding by generative pre-training.

\bibitem[{Radford et~al.(2019)Radford, Wu, Child, Luan, Amodei, Sutskever et~al.}]{radford2019language}
Alec Radford, Jeffrey Wu, Rewon Child, David Luan, Dario Amodei, Ilya Sutskever, et~al. 2019.
\newblock Language models are unsupervised multitask learners.
\newblock \emph{OpenAI blog}, 1(8):9.

\bibitem[{Raffel et~al.(2023)Raffel, Shazeer, Roberts, Lee, Narang, Matena, Zhou, Li, and Liu}]{raffel2023exploringlimitstransferlearning}
Colin Raffel, Noam Shazeer, Adam Roberts, Katherine Lee, Sharan Narang, Michael Matena, Yanqi Zhou, Wei Li, and Peter~J. Liu. 2023.
\newblock \href {http://arxiv.org/abs/1910.10683} {Exploring the limits of transfer learning with a unified text-to-text transformer}.

\bibitem[{Sch{\"o}nemann(1966)}]{schonemann1966generalized}
Peter~H Sch{\"o}nemann. 1966.
\newblock A generalized solution of the orthogonal procrustes problem.
\newblock \emph{Psychometrika}, 31(1):1--10.

\bibitem[{Schuster et~al.(2019)Schuster, Ram, Barzilay, and Globerson}]{schuster-etal-2019-cross}
Tal Schuster, Ori Ram, Regina Barzilay, and Amir Globerson. 2019.
\newblock \href {https://doi.org/10.18653/v1/N19-1162} {Cross-lingual alignment of contextual word embeddings, with applications to zero-shot dependency parsing}.
\newblock In \emph{Proceedings of the 2019 Conference of the North {A}merican Chapter of the Association for Computational Linguistics: Human Language Technologies, Volume 1 (Long and Short Papers)}, pages 1599--1613, Minneapolis, Minnesota. Association for Computational Linguistics.

\bibitem[{Shetty et~al.(2024)Shetty, Xu, and Lau}]{shetty2024wet}
Anudeex Shetty, Qiongkai Xu, and Jey~Han Lau. 2024.
\newblock Wet: Overcoming paraphrasing vulnerabilities in embeddings-as-a-service with linear transformation watermarks.
\newblock \emph{arXiv preprint arXiv:2409.04459}.

\bibitem[{Smith et~al.(2017)Smith, Turban, Hamblin, and Hammerla}]{smith2017offline}
Samuel~L Smith, David~HP Turban, Steven Hamblin, and Nils~Y Hammerla. 2017.
\newblock Offline bilingual word vectors, orthogonal transformations and the inverted softmax.
\newblock \emph{arXiv preprint arXiv:1702.03859}.

\bibitem[{Socher et~al.(2013)Socher, Perelygin, Wu, Chuang, Manning, Ng, and Potts}]{socher-etal-2013-recursive}
Richard Socher, Alex Perelygin, Jean Wu, Jason Chuang, Christopher~D. Manning, Andrew Ng, and Christopher Potts. 2013.
\newblock \href {https://aclanthology.org/D13-1170} {Recursive deep models for semantic compositionality over a sentiment treebank}.
\newblock In \emph{Proceedings of the 2013 Conference on Empirical Methods in Natural Language Processing}, pages 1631--1642, Seattle, Washington, USA. Association for Computational Linguistics.

\bibitem[{Song and Raghunathan(2020)}]{10.1145/3372297.3417270}
Congzheng Song and Ananth Raghunathan. 2020.
\newblock \href {https://doi.org/10.1145/3372297.3417270} {Information leakage in embedding models}.
\newblock In \emph{Proceedings of the 2020 ACM SIGSAC Conference on Computer and Communications Security}, CCS '20, page 377–390, New York, NY, USA. Association for Computing Machinery.

\bibitem[{Wang et~al.(2019{\natexlab{a}})Wang, Xiong, Yu, Guo, Chang, and Wang}]{wang-etal-2019-sentence}
Hong Wang, Wenhan Xiong, Mo~Yu, Xiaoxiao Guo, Shiyu Chang, and William~Yang Wang. 2019{\natexlab{a}}.
\newblock \href {https://doi.org/10.18653/v1/N19-1086} {Sentence embedding alignment for lifelong relation extraction}.
\newblock In \emph{Proceedings of the 2019 Conference of the North {A}merican Chapter of the Association for Computational Linguistics: Human Language Technologies, Volume 1 (Long and Short Papers)}, pages 796--806, Minneapolis, Minnesota. Association for Computational Linguistics.

\bibitem[{Wang et~al.(2019{\natexlab{b}})Wang, Che, Guo, Liu, and Liu}]{wang-etal-2019-cross}
Yuxuan Wang, Wanxiang Che, Jiang Guo, Yijia Liu, and Ting Liu. 2019{\natexlab{b}}.
\newblock \href {https://doi.org/10.18653/v1/D19-1575} {Cross-lingual {BERT} transformation for zero-shot dependency parsing}.
\newblock In \emph{Proceedings of the 2019 Conference on Empirical Methods in Natural Language Processing and the 9th International Joint Conference on Natural Language Processing (EMNLP-IJCNLP)}, pages 5721--5727, Hong Kong, China. Association for Computational Linguistics.

\bibitem[{Wiesinger et~al.(2025)Wiesinger, Marlow, and Vuskovic}]{agents_google}
Julia Wiesinger, Patrick Marlow, and Vladimir Vuskovic. 2025.
\newblock \href {https://www.kaggle.com/whitepaper-agents} {Agents}.
\newblock Accessed: 2025-01-07.

\bibitem[{Xue et~al.(2021)Xue, Constant, Roberts, Kale, Al-Rfou, Siddhant, Barua, and Raffel}]{xue-etal-2021-mt5}
Linting Xue, Noah Constant, Adam Roberts, Mihir Kale, Rami Al-Rfou, Aditya Siddhant, Aditya Barua, and Colin Raffel. 2021.
\newblock \href {https://doi.org/10.18653/v1/2021.naacl-main.41} {m{T}5: A massively multilingual pre-trained text-to-text transformer}.
\newblock In \emph{Proceedings of the 2021 Conference of the North American Chapter of the Association for Computational Linguistics: Human Language Technologies}, pages 483--498, Online. Association for Computational Linguistics.

\end{thebibliography}
\bibliographystyle{acl_natbib}

\appendix

\newpage
\section{Derivation of Normal Equation}~\label{normal_equation}
To determine the optimal transformation matrix $\mW$, we aim to minimize a cost function $J$ that quantifies the discrepancy between the attack embedding matrix $\mE_{A}$ and the victim embeddings $\mE_{V}$: 

\begin{equation}
\begin{aligned}
J(\mW) &= \frac{1}{2} (\mE_A  - \mE_V \mW)^{T} (\mE_A - \mE_V \mW) \\
& = \frac{1}{2}(\mE_A^{T} \mE_A - \mE_A^{T} \mE_V \mW -  (\mE_{V} \mW)^{T} \mE_{A} \\
& + (\mE_{V} \mW)^{T} \mE_V \mW) \\
& = \frac{1}{2}(\mE_A^{T} \mE_A - \mE_A^{T} \mE_V \mW -   \mW^{T}\mE_{V}^{T} \mE_{A} \\
& +  \mW^{T}\mE_{V}^{T} \mE_V \mW)
\end{aligned}
\end{equation}

To compute the derivatives of $J(\mW)$:

\begin{equation}
    \begin{aligned}
        \nabla_{\mW} J(\mW)  & =\frac{1}{2} \nabla_{\mW} (\mE_A^{T} \mE_A - \mE_A^{T} \mE_V \mW \\
        & - \mW^{T}\mE_{V}^{T} \mE_{A} +  \mW^{T}\mE_{V}^{T} \mE_V \mW) \\
        & = 2\mE_{V}^{T} \mE_V \mW -2 \mE_{V}^{T} \mE_{A} .
    \end{aligned}
\end{equation}

To minimize $J$, setting its derivatives to 0, we obtain the normal equation :
\begin{equation}
    \mE^{T}_{V} \mE_{V} \mW = \mE^{T}_{V} \mE_{A}.
\end{equation}

The matrix $\mW$ that minimizes $J(\mW)$ is

\begin{equation}
\mW = (\mE_V^{T}\mE_V)^{-1}\mE^{T}_{V} \mE_{A}.
\end{equation}

\begin{table*}[t!]
    \centering
    \resizebox{\linewidth}{!}{
    \begin{tabular}{l|l|l|l|l}
    \toprule
    Model  &  Huggingface & Architecture & \#Languages &   Dimension\\
    \midrule
       \textsc{Flan-T5}~\citep{chung2022scalinginstructionfinetunedlanguagemodels}  & google/flan-t5-small  & Encoder-Decoder  & 60  & 512 \\
        ~\textsc{GTR}~\citep{ni2021largedualencodersgeneralizable} & sentence-transformers/gtr-t5-base &  Encoder & 1  & 768\\
    ~\textsc{T5}~\citep{raffel2023exploringlimitstransferlearning} & google-t5/t5-base &  Encoder-Decoder & 4 & 768\\
      ~\textsc{mT5}~\citep{xue-etal-2021-mt5} & google/mt5-base &   Encoder-Decoder & 102  &768\\
    ~\textsc{mBERT}~\citep{devlin2019bertpretrainingdeepbidirectional} & google-bert/bert-base-multilingual-cased &  Encoder  &  104 &768 \\
    ~\textsc{text-embedding-ada-002}& OpenAI API &  Encoder & 100+ & 1536 \\
    ~\textsc{text-embedding-3-large} & OpenAI API & Encoder & 100+ & 3072 \\
    \bottomrule  
    \end{tabular}}
    \caption{Details of LLMs and Embeddings.}
    \label{tab:llms}
\end{table*}


\section{Defense Mechanisms}

\subsection{WET}~\label{wet_algo}


$\mT$ is constructed by adopting circulant matrices~\citep{gray2006toeplitz} to ensure that the transformation matrix is both full-rank and well-conditioned to allow for accurate pseudoinverse computation for recovering the original embeddings from watermarked embeddings~\citep{shetty2024wet}, refer to~\citet{shetty2024wet} for the complete algorithm for generating $\mT$.

In detail, WET as a defense is applied to aligned embeddings with the equation~\ref{eq:wet_equation_tranformation}, where $\mW$ is the optimal solution for alignment, and $\mT$ is invertible.

\begin{equation}\label{eq:wet_equation_tranformation}
    \begin{aligned}
    \text{Norm}(\mT \mE_{V \to A} )
    & =  \text{Norm}(\mT  (\mE_V  \mW))\\
    & = \text{Norm}((\mT\mE_V)\mW )
    \end{aligned}
\end{equation}

\subsection{(Local) Differential Privacy (DP)}~\label{ldp_appendix}

As illustrated in~\citet{du2023sanitizing}, DP ensures that a randomized mechanism $\gM$ behaves similarly on any two neighboring datasets $\gX\simeq \gX'$ differing in only one individual's contribution (e.g., a sequence).
It is formally defined as follows:

\begin{definition}
    Let $\epsilon \geq 0$, $0\leq \delta \leq 1$ be \textit{two privacy parameters}. $\gM$ fulfills $(\epsilon, \delta)-$DP, if $\forall \gX\simeq \gX'$ and any output set $\gO \subseteq Range(\gM) $, 
    $\Pr[\gM(\gX)\in \gO]\leq e^{\epsilon} \cdot \Pr[\gM(\gX')\in \gO] + \delta$.
\end{definition}
If $\delta=0$, then we say that $\gM$ is $\epsilon-$DP or pure DP.

There are two popular DP settings, \textit{central} and \textit{local}.
In central DP, a trusted curator can access the raw data of all individuals, apply a Mechanism $\gM$ with random noise to ensure DP, and then release the perturbed outputs.
Local DP (LDP) is ensured without the curator by letting individuals perturb their data locally before being shared. The local DP~\citep{kasiviswanathan2011can} is defined as follows:

\begin{definition}
    Let $\epsilon \geq 0$ be a \textit{privacy parameter}. $\gM$ is $\epsilon-$LDP, 
    if for any two private inputs $\gX$, $\gX'$ and \textit{any output set} $\gO \subseteq Range(\gM)$,
    $\Pr[\gM(\mX) \in \gO] \leq e^{\epsilon} \cdot \Pr[\gM(\gX')\in \gO]$.
    
\end{definition}
However, $\epsilon-LDP$ offers homogenous protection for all input pairs, which can be too stringent in certain scenarios. When $\epsilon$ is too small, the noisy outputs are useless for utility tasks.

Thus,~\citet{du2023sanitizing} customizes heterogeneous privacy guarantees for different pairs of inputs, so called metric-based LDP, formally defined as follows:

\begin{definition}
    Let $\epsilon\geq 0 $ be the \textit{privacy parameter}, and $d$ be a suitable distance metric for the input space. 
    $\gM$ satisfies $\epsilon d$-LDP, if for any two inputs $\gX$, $\gX'$ and any output set $\gO \subseteq Range(\gM)$, 
    $\Pr[\gM(\mX) \in \gO] \leq e^{\epsilon d(\mX, \mX')} \cdot \Pr[\gM(\mX')\in \gO]$.
\end{definition}

\textit{Purkayastha Mechanism} with Purkayastha distribution and \textit{Planar Lapalace Mechanism} with Euclidean metric, are thus proposed to ensure $\epsilon d$-LDP on embeddings. Refer to~\citet{du2023sanitizing} for details in transforming embeddings with these mechanisms.

\section{Ablation Study of Leakage Data Size}\label{apendix:datasize}

\begin{figure*}[t!]
    \centering
    \includegraphics[width=\linewidth]{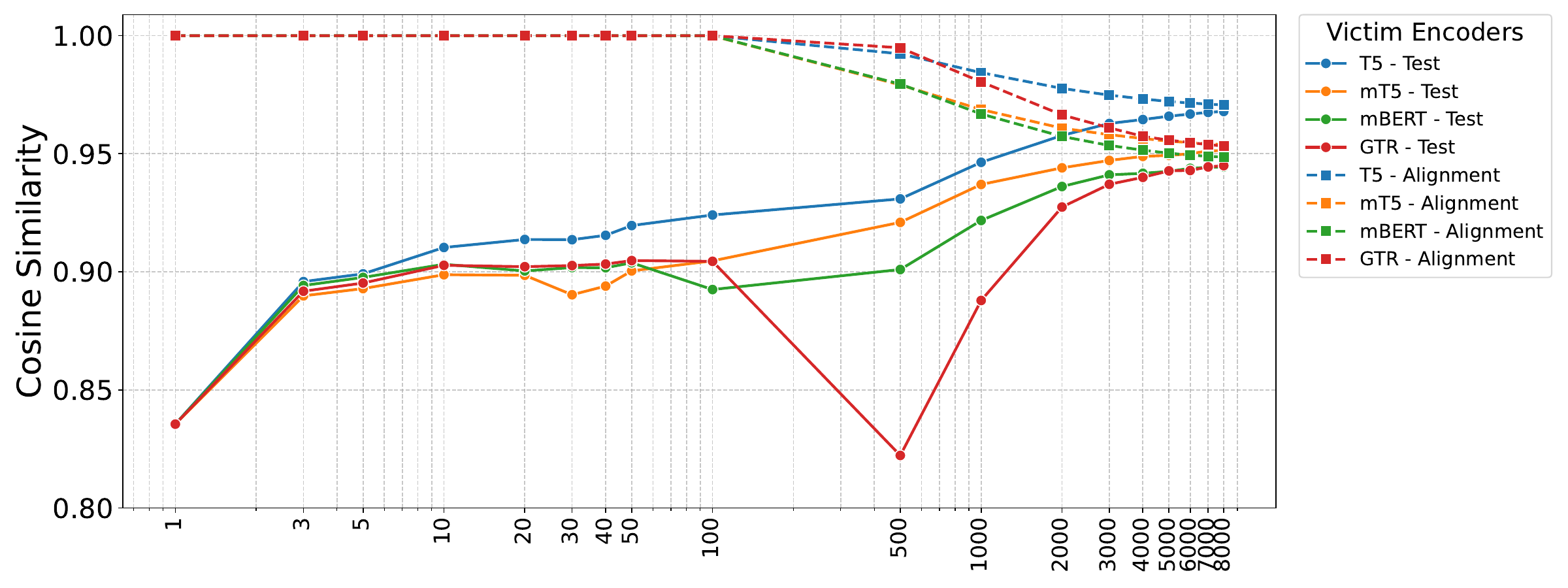}
    \includegraphics[width=0.95\linewidth]{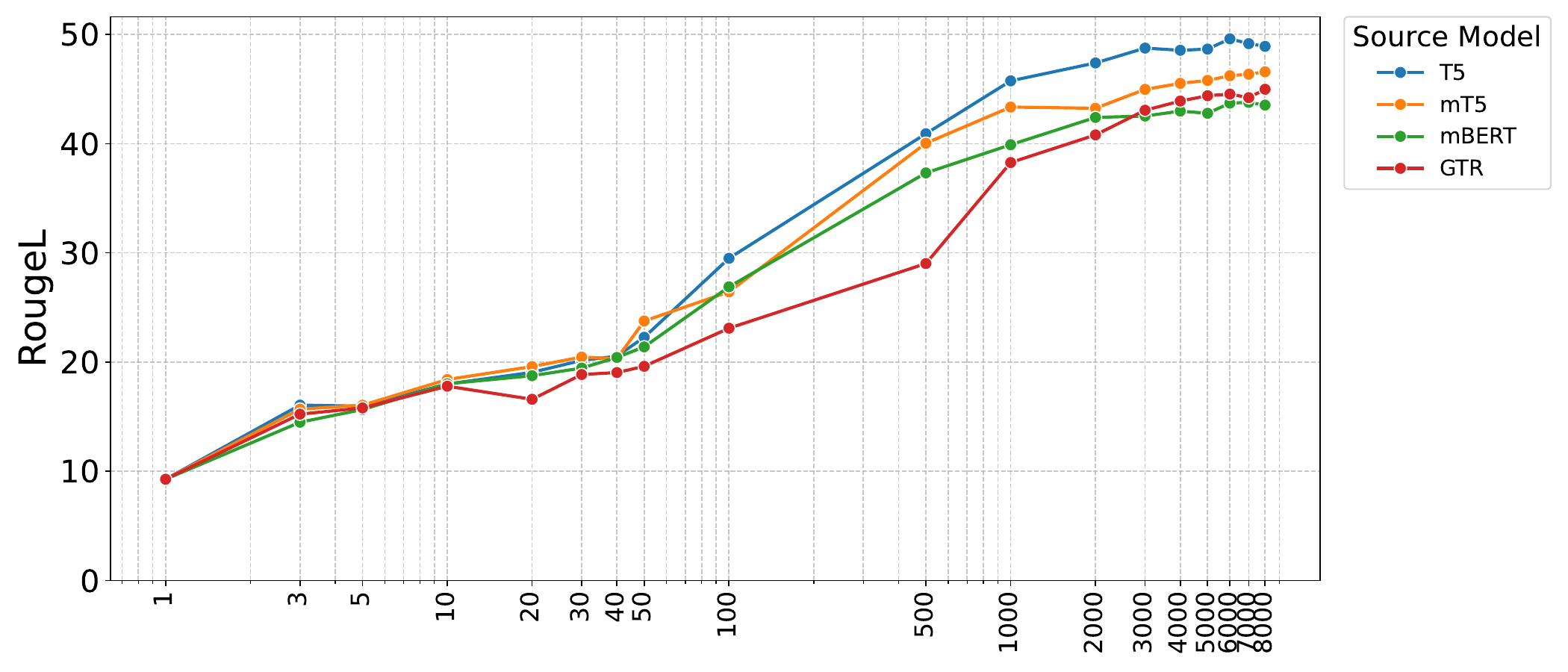}
    \caption{Inversion Performance vs. Leakage Data (Alignment) Sizes. (Top) shows the cosine similarities for embeddings of alignment data (dashed) and for embeddings of test data by the size of alignment data. (Bottom) shows the Rouge-L scores for the }
    \label{fig:datasize_ablation}
\end{figure*}

The more data for alignment, the better the performance for \textbf{\ourmethod}. However, after a certain amount of data, i.e., 3K, the increase in data samples does not boost the inversion performance.
We conduct an ablation study of the sizes of leakage data for alignment in terms of inversion performances.
Fig.~\ref{fig:datasize_ablation} shows the inversion performance in Cosine Similarity (Top) and Rouge-L (Bottom) with the leakage data sizes from 1 to 8k.
As shown in Fig.~\ref{fig:datasize_ablation} (Top), embeddings for alignment are the perfect match for cosine similarities from 1 to 100 samples, then they decrease until they converge with the cosine similarities for aligned test embeddings for the corresponding encoder.
In Fig.~\ref{fig:datasize_ablation} (Top), while the inversion performances in Rouge-L increase with more data points for alignment until 8k across the encoders, the performance increases sharply from 1 to 1k. It becomes relatively stagnant from 2k to 8k.
Considering the trade-off between inversion performance and the size of data samples, we choose 1k as the upper bound of the number of data samples for alignment to conduct thorough experimentation in this work.


\begin{figure}[htbp]
    \centering

    \includegraphics[width=\linewidth]{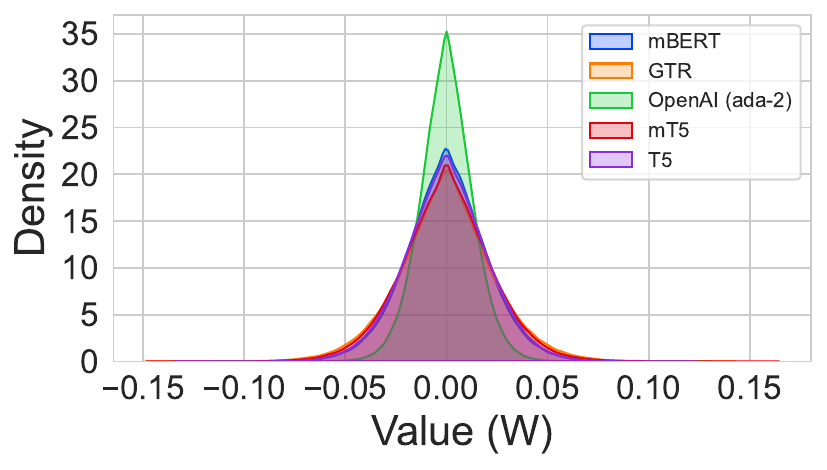}
    (a) SNLI
    \includegraphics[width=\linewidth]{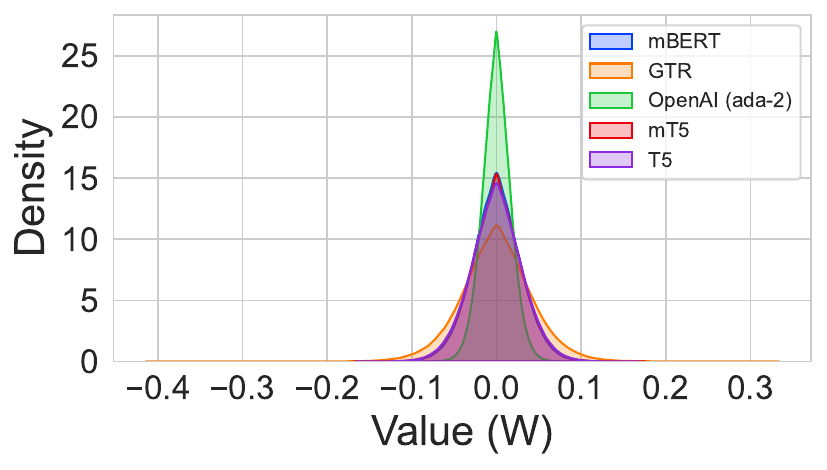} 
    (b) SST2
    \includegraphics[width=\linewidth]{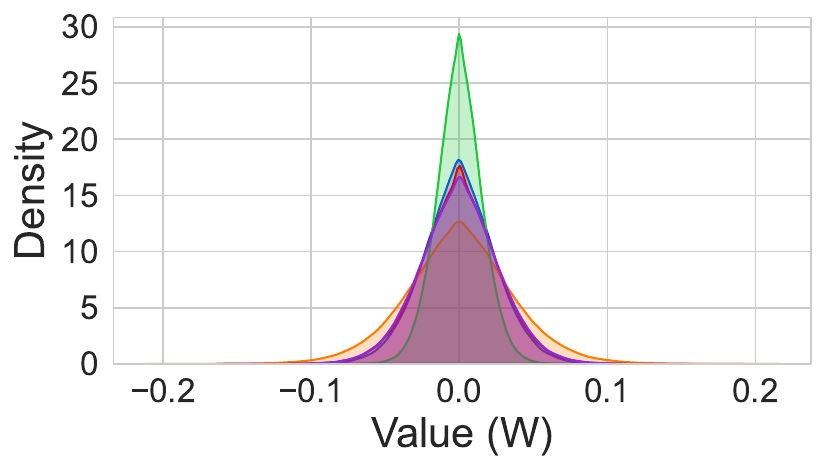}
    (C) S140
    \caption{The Analysis of Alignment Transformation Weight ($\mW$) on Victim Encoders on different datasets.}
    \label{fig:W_alignment_density}
\end{figure}

\begin{table*}[!ht]
    \centering
    \resizebox{\linewidth}{!}{
    \begin{tabular}{l|p{14cm}}
    \hline
    \toprule
            Input & This business uses tools provided by \textcolor{teal}{\textbf{TripAdvisor [ORG]}}(or one of its official Review Collection Partners) to encourage and collect guest reviews, including this one.  \\ 

        Reconstructed & This business uses tools provided by \textcolor{red}{\textbf{TripAdvisor [ORG]}} (or one of its official Review Collection Partners) to encourage and collect guest reviews, including this one \\
        \midrule
                Input & Book your flights now from  Hermosillo (\textcolor{teal}{\textbf{Mexico [GPE]}}) to the most important cities in the world. The box below contains flights from Her. \\

        Reconstructed & Book your flights now from  \colorbox{lightgray}{Las Vegas} (\textcolor{red}{\textbf{Mexico [GPE]}}) to the most important cities in the world. The box below contains flights from \colorbox{lightgray}{Las Vegas to} \\
                \midrule

                Input & KOffice Project Reviews Starts: \textcolor{teal}{\textbf{1, 2, 3, 4, 5 [CARDINAL]}} with comment only In chronological order from new to old - KOffice - OSDN Download. \\

        Reconstructed & \colorbox{lightgray}{DevOps} Project Reviews Starts: \textcolor{red}{\textbf{1, 2, 3, 4, 5 [CARDINAL]}} with comment only In chronological order from new to old - \colorbox{lightgray}{DevOp}  \\

                       \midrule

        Input & \textcolor{teal}{\textbf{TripAdvisor [ORG]}} is proud to partner with \textcolor{teal}{\textbf{Travelocity, Expedia, Hotels.com, Agoda, Booking.com, Price [ORG]}}. \\

        Reconstructed & \textcolor{red}{\textbf{TripAdvisor [ORG]}} is proud to partner with \textcolor{red}{\textbf{Booking.com, Expedia, Hotels.com, Travelocity, Agoda, Price [ORG]}} \\ 

         \midrule
         Input & Step \textcolor{teal}{\textbf{5 [CARDINAL]}}: Utilize Windows System Restore to """"Undo"""" Recent System Changes Windows System Restore allows you to """"go back in. \\
         
        Reconstructed & Step \textcolor{red}{\textbf{5 [CARDINAL]}}: Utilize Windows System Restore to """"Undo"""" Recent System Changes Windows System Restore allows you to """"go back in \\

                \midrule

        Input & If you want to ensure you grab a bargain, try to book \textcolor{teal}{\textbf{more than 90 days [DATE]}} before your stay to get the best price for a \textcolor{teal}{\textbf{Paris [GPE]}}.\\
         Reconstructed & If you want to ensure you grab a bargain, try to book \textcolor{red}{\textbf{more than 90 days [DATE]}} before your stay to get the best price for a \textcolor{red}{\textbf{Paris [GPE]}} \\
                       \midrule
        Input  & The Fund’s total amount for the Fund is limited to a maximum of \textcolor{red}{\textbf{\$4,000,000 [MONEY]}}, and the Fund’s total amount for each transaction is\\

        Reconstructed & \colorbox{lightgray}{means an amount up to} \textcolor{teal}{\textbf{USD 4,000,000 [MONEY]}}, \colorbox{lightgray}{for the Winners, the exact amount } \colorbox{lightgray}{is subject to the sole and final discretion of} the Fund;\\
                        \midrule
        Input & \textcolor{teal}{\textbf{Microsoft [ORG]}} is constantly updating and improving \textcolor{teal}{\textbf{Windows [PRODUCT]}} system files that could be associated with 100street\_bkg\_bikini\_bottom.swf.\\
        Reconstructed & \textcolor{red}{\textbf{Microsoft [ORG]}} is constantly updating and improving \textcolor{red}{\textbf{Windows [PRODUCT]}} system files that could be associated with \colorbox{lightgray}{jabber-shp-src.jar. Sometimes} \\

        \bottomrule

    \end{tabular}
    }
    \caption{Qualitative Analysis of In-Domain Inversion Results from \textsc{OpenAI (ada-2)} embeddings with 1k alignment data samples and the attack model trained on MultiPHLT English dataset. The matched entities with their entity types are colored and \textbf{bolded} in \textcolor{teal}{\textbf{Input}} and \textcolor{red}{\textbf{Reconstructed}}. The mismatched reconstructed texts are in \colorbox{lightgray}{grey colored box}. 
    [GPE]: Countries/cities/states; [ORG]:Organization.
    }
    \label{tab:qualitative}
\end{table*}

\begin{table}[htbp]
    \centering
    \resizebox{\linewidth}{!}{
    \begin{tabular}{c|c|cc|cc}
    \toprule 
        \textbf{Victim} & \textbf{Defense} & \textbf{Rouge-L}$\downarrow$ & \textbf{COS}$\downarrow$ & \textbf{ACC}$\uparrow$ & \textbf{F1}$\uparrow$ \\ 
        \midrule
        \textsc{random} &- &  6.57 & -0.0108 & 56.50 & 56.02 \\

        \midrule 
          & - & 25.06 & 0.9359 & 89.5 & 89.5 \\ 
         \cmidrule{2-6} 
          & WET & 24.52 & 0.9344 & 87 & 87 \\ 
         \cmidrule{2-6} 
          & Shuffling & 20.34 & 0.9358 & 89 & 89 \\ 
         \cmidrule{2-6} 
        \textsc{T5} & 0.001 & 25.15 & 0.9355 & 89.5 & 89.5 \\ 
          & 0.005 & 24.59 & 0.9308 & 88.5 & 88.5 \\ 
          & 0.01 & 22.06 & 0.9172 & 88 & 88 \\ 
          & 0.05 & 12.11 & 0.7637 & 87.5 & 87.5 \\ 
          & 0.1 & 9.57 & 0.7092 & 79 & 79 \\ 
          & 0.5 & 7.03 & 0.3169 & 59 & 59 \\ 
          & 1 & 7.05 & 0.1878 & 56.5 & 56.41 \\ 
        \midrule
          & - & 18.14 & 0.8823 & 85.5 & 85.5 \\ 
         \cmidrule{2-6} 
          & WET & 15.69 & 0.9178 & 84 & 83.97 \\ 
         \cmidrule{2-6} 
          & Shuffling & 14.69 & 0.8835 & 85.5 & 85.5 \\ 
         \cmidrule{2-6} 
        \textsc{GTR} & 0.001 & 17.22 & 0.8804 & 85 & 84.99 \\ 
          & 0.005 & 17.01 & 0.8748 & 85 & 84.99 \\ 
          & 0.01 & 15.09 & 0.8489 & 85 & 84.99 \\ 
          & 0.05 & 9.91 & 0.7136 & 83.5 & 83.5 \\ 
          & 0.1 & 9.18 & 0.6505 & 76 & 75.91 \\ 
          & 0.5 & 7.53 & 0.2297 & 55 & 54.93 \\ 
          & 1 & 6.62 & 0.0293 & 49 & 48.87 \\ 
        \midrule
          & - & 21.83 & 0.9320 & 79.5 & 79.47 \\ 
         \cmidrule{2-6} 
          & WET & 21.04 & 0.9307 & 80 & 79.93 \\ 
         \cmidrule{2-6} 
          & Shuffling & 18.19 & 0.9321 & 78.5 & 78.47 \\ 
         \cmidrule{2-6} 
        \textsc{mT5} & 0.001 & 22.13 & 0.9327 & 77.5 & 77.43 \\ 
          & 0.005 & 21.71 & 0.9277 & 80 & 79.99 \\ 
         & 0.01 & 18.97 & 0.9119 & 79.5 & 79.41 \\ 
          & 0.05 & 9.72 & 0.7410 & 77.5 & 77.34 \\ 
          & 0.1 & 8.83 & 0.6924 & 64 & 62.79 \\ 
          & 0.5 & 7.55 & 0.2407 & 51 & 50.98 \\ 
          & 1 & 6.55 & 0.1930 & 49 & 48.95 \\ 
        \midrule
          & - & 20.44 & 0.9211 & 76.5 & 76.06 \\ 
         \cmidrule{2-6} 
          & WET & 20.19 & 0.9261 & 77 & 76.66 \\ 
         \cmidrule{2-6} 
          & Shuffling & 17.07 & 0.9209 & 76 & 75.59 \\ 
         \cmidrule{2-6} 
        \textsc{mBERT} & 0.001 & 20.32 & 0.9209 & 76.5 & 76.06 \\ 
          & 0.005 & 20.01 & 0.9156 & 77 & 76.6 \\ 
          & 0.01 & 18.22 & 0.9018 & 76 & 75.59 \\ 
          & 0.05 & 10.97 & 0.7320 & 69.5 & 68.54 \\ 
          & 0.1 & 8.6 & 0.6851 & 64.5 & 64.02 \\ 
          & 0.5 & 7.06 & 0.3407 & 59.5 & 59.5 \\ 
          & 1 & 6.12 & 0.0838 & 46 & 45.86 \\ 
        \midrule
       & - & 20.15 & 0.9309 & 92 & 92 \\ 
         \cmidrule{2-6} 
          & WET & 15.83 & 0.9258 & 91 & 91 \\ 
         \cmidrule{2-6} 
         & Shuffling & 17.74 & 0.9311 & 91.5 & 91.5 \\ 
         \cmidrule{2-6} 
        \shortstack{\textsc{OpenAI}\\ \textsc{(ada-2)}}
        
        & 0.001 & 20.11 & 0.9308 & 91 & 91 \\ 
         & 0.005 & 20.01 & 0.9300 & 91 & 91 \\ 
         & 0.01 & 17.57 & 0.9179 & 90 & 90 \\ 
         & 0.05 & 9.39 & 0.7787 & 82.5 & 82.5 \\ 
        & 0.1 & 8.07 & 0.7502 & 71 & 70.9 \\ 
        & 0.5 & 7.81 & 0.4743 & 51.5 & 51.4 \\ 
          & 1 & 5.92 & 0.1324 & 53.5 & 52.2 \\ 
        \bottomrule
    \end{tabular}}
     \caption{The Inversion and Utility Performance on Classification Tasks on SST2 dataset with WET, Shuffling, Guassian Noise Insertion. From a defender's perspective, $\uparrow$ means higher are better, $\downarrow$ means lower are better.}
     \label{tab:wet_shuffling_gaussian_sst2}
\end{table}

\begin{table}[htbp]
     \centering
    \resizebox{\linewidth}{!}{
    \begin{tabular}{c|c|cc|cc}
    \toprule 
        \textbf{Victim} & \textbf{Defense} & \textbf{Rouge-L}$\downarrow$ & \textbf{COS}$\downarrow$ & \textbf{ACC}$\uparrow$ & \textbf{F1}$\uparrow$ \\ 
        \midrule 

       \textsc{random} & - & 5.01  & 0.0275  & 48.5  & 48.13  \\

        \midrule
          & - & 23.04 & 0.9219 & 71.5 & 70.37 \\ 
          \cmidrule{2-6} 
          & WET & 19.86 & 0.9186 & 69 & 67.42 \\ 
          \cmidrule{2-6} 
          & Shuffling & 17.56 & 0.9245  & 71 & 70.03 \\ 
          \cmidrule{2-6} 
         \textsc{T5} & 0.001 & 22.05 & 0.9230 & 71.5 & 70.37 \\ 
         & 0.005 & 21.11 & 0.9188 & 70.5 & 69.45 \\ 
          & 0.01 & 18.29 & 0.9078 & 69 & 67.84 \\ 
          & 0.05 & 8.85 & 0.7561 & 68 & 66.8 \\ 
          & 0.1 & 7.6 & 0.6901 & 59.5 & 58.86 \\ 
          & 0.5 & 4.97 & 0.2678 & 46 & 45.95 \\ 
          & 1 & 4.85 & 0.1101 & 50.5 & 50.47 \\ 
          \midrule
          & - & 13.22 & 0.8585 & 70.5 & 68.45 \\
        \cmidrule{2-6} 
          & WET & 12.15 & 0.8861 & 71.5 & 69.35 \\ 
        \cmidrule{2-6} 
          & Shuffling & 11.69 & 0.8599 & 70 & 68 \\ 
        \cmidrule{2-6} 
        \textsc{GTR} & 0.001 & 13.59 & 0.8594 & 69.5 & 67.38 \\ 
          & 0.005 & 12.86 & 0.8515 & 70 & 68.17 \\ 
          & 0.01 & 12.11 & 0.8260 & 71 & 69.23 \\ 
          & 0.05 & 7.92 & 0.7055 & 70.5 & 69.21 \\ 
          & 0.1 & 5.8 & 0.6137 & 64 & 63.28 \\ 
          & 0.5 & 4.6 & 0.3075 & 56.5 & 56.47 \\ 
          & 1 & 4.23 & 0.0582 & 53 & 52.83 \\ 
        \midrule
          & - & 19.82 & 0.9212  & 68 & 66.8 \\ 
                  \cmidrule{2-6} 

          & WET & 18.4 & 0.9219  & 63 & 61.61 \\ 
                  \cmidrule{2-6} 
          & Shuffling & 16.51 & 0.9215  & 66.5 & 65.44 \\ 
                  \cmidrule{2-6} 
        \textsc{mT5} & 0.001 & 20.4 & 0.9217 & 67.5 & 66.47 \\ 
          & 0.005 & 18.82 & 0.9173 & 67 & 65.89 \\ 
          & 0.01 & 16.48 & 0.9037 & 65.5 & 64.28 \\ 
          & 0.05 & 8.09 & 0.7625 & 58.5 & 57.03 \\ 
          & 0.1 & 7.02 & 0.6952 & 60 & 59.6 \\ 
          & 0.5 & 5.13 & 0.3770 & 55.5 & 55.31 \\ 
          & 1 & 4.21 & 0.1505 & 42 & 41.99 \\ 
          \midrule
          & - & 17.88 & 0.9070 & 64.5 & 64.18 \\
          \cmidrule{2-6} 

          & WET & 16.83 & 0.9151  & 63 & 62.63 \\ 
          \cmidrule{2-6} 
          & Shuffling & 13.8 & 0.9062 & 63.5 & 63.29 \\ 
          \cmidrule{2-6} 
        mBERT & 0.001 & 17.83 & 0.9077 & 64 & 63.77 \\ 
          & 0.005 & 17.01 & 0.9020 & 64.5 & 64.24 \\ 
          & 0.01 & 15.44 & 0.8901 & 63.5 & 63.17 \\ 
          & 0.05 & 8.47 & 0.7498 & 60.5 & 60.21 \\ 
          & 0.1 & 7.26 & 0.6876 & 52 & 50.39 \\ 
          & 0.5 & 5.25 & 0.2904 & 50 & 49.68 \\ 
          & 1 & 4.34 & 0.1619 & 55.5 & 55.31 \\ 
          \midrule
       & - & 17.13 & 0.9224 & 71.5 & 70.12 \\ 
        \cmidrule{2-6} 
         & WET & 14.19 & 0.9237 & 69.5 & 67.55 \\ 
         \cmidrule{2-6} 
        & Shuffling & 15.97 & 0.9229 & 71 & 69.53 \\ 
        \cmidrule{2-6} 
        \shortstack{\textsc{OpenAI}\\ \textsc{(ada-2)}} & 0.001 & 17.12 & 0.9225 & 71 & 69.53 \\ 
     & 0.005 & 17.19 & 0.9208 & 72.5 & 71.17 \\ 
          & 0.01 & 14.8 & 0.9100 & 71.5 & 70.12 \\ 
        & 0.05 & 7.4 & 0.7974 & 74 & 73.13 \\ 
         & 0.1 & 6.42 & 0.7691 & 67.5 & 67.4 \\ 
        & 0.5 & 5.36 & 0.4271 & 53.5 & 53.16 \\ 
         & 1 & 5.27 & 0.1753 & 55.5 & 54.09 \\ 
        \bottomrule
    \end{tabular}}
    \caption{The Inversion and Utility Performance on Classification Tasks on S140 dataset with WET, Shuffling, Guassian Noise Insertion. From a defender's perspective, $\uparrow$ means higher are better, $\downarrow$ means lower are better.}
         \label{tab:wet_shuffling_gaussian_s140}

\end{table}

\begin{table*}[htbp]
    \centering
     \resizebox{0.65\linewidth}{!}{
    \begin{tabular}{l|l|cccc|cccc}
            \toprule

        Victim Model & $\epsilon$ & Rouge-L & COS & ACC & F1 & Rouge-L & COS & ACC & F1 \\ 
        \midrule
        ~ & &  \multicolumn{4}{c|}{LapMech} &  \multicolumn{4}{c}{PurMech} \\ 
         \midrule
          & - & 25.06 & 0.9359 & 89.5 & 89.5 & 23.04 & 0.9219 & 71.5 & 70.37 \\ 
      \cmidrule{2-10} 
          & 1 & 6.84 & 0.0103 & 54.5 & 54.5 & 4.19 & -0.0235 & 55 & 54.93 \\ 
        \textsc{T5} & 4 & 5.94 & -0.0660 & 76 & 75.96 & 4.48 & 0.0322 & 63.5 & 63.39 \\ 
    
          & 8 & 6.53 & 0.0526 & 84.5 & 84.5 & 4.29 & -0.1135 & 71 & 70.03 \\ 
          & 12 & 6.78 & 0.0551 & 89.5 & 89.49 & 4.29 & -0.0210 & 71 & 70.03 \\
          \midrule
          & - & 18.14 & 0.8823 & 85.5 & 85.5 & 13.22 & 0.8585 & 70.5 & 68.45   \\       \cmidrule{2-10} 

          & 1 & 5.86 & -0.0213 & 55.5 & 55.5 & 4.29 & -0.0079 & 50 & 49.82 \\ 
        \textsc{GTR} & 4 & 6.64 & 0.0232 & 73 & 72.98 & 4.55 & -0.0066 & 61 & 58.4 \\ 
          & 8 & 6.75 & 0.0172 & 83 & 82.99 & 4.82 & -0.0179 & 67 & 65.16 \\ 
          & 12 & 7.05 & 0.0157 & 84 & 84 & 4.85 & 0.0439 & 70 & 68.75 \\ 
          \midrule
          & - & 21.83 & 0.9320 & 79.5 & 79.47 & 17.13 & 0.9224 & 71.5 & 70.12 \\ 
        \cmidrule{2-10} 
        \textsc{mT5} & 1 & 7.18 & -0.0850 & 54 & 53.98 & 3.89 & -0.0560 & 55 & 54.93 \\ 
          & 4 & 6.78 & 0.0471 & 69.5 & 69.49 & 4.09 & -0.0255 & 66.5 & 65.03 \\ 
          & 8 & 6.75 & -0.0008 & 78 & 78 & 4.9 & -0.0166 & 68.5 & 66.66 \\ 
          & 12 & 6.71 & 0.0638 & 81 & 80.95 & 5.3 & 0.1235 & 69.5 & 67.72 \\ 
          
          \midrule
         & - & 20.44 & 0.9211 & 76.5 & 76.06 & 17.88 & 0.9070 & 64.5 & 64.18 \\ 
         \cmidrule{2-10} 
        \textsc{mBERT} & 1 & 6.59 & 0.0182 & 50.5 & 50.5 & 4.23 & 0.0215 & 49.5 & 49.5 \\ 
          & 4 & 6.11 & -0.1486 & 70 & 69.97 & 4.6 & -0.0190 & 63.5 & 63.46 \\ 
          & 8 & 6.76 & -0.0680 & 76 & 75.8 & 4.26 & 0.0942 & 63 & 62.91 \\ 
          & 12 & 6.55 & 0.0431 & 78 & 77.68 & 4.66 & -0.0340 & 62 & 61.69 \\

          \midrule
             & - & 20.15 & 0.9309 & 92 & 92 & 19.82 & 0.9212 & 68 & 66.8 \\
                 \cmidrule{2-10} 

         & 1 & 6.06 & -0.0289 & 50.5 & 50.47 & 4.52 & 0.0507 & 53 & 52.98 \\ 
        \shortstack{\textsc{OpenAI}\\ \textsc{(ada-2)}}   & 4 & 6.56 & 0.0336 & 79 & 79 & 4.14 & -0.0377 & 63.5 & 62.6 \\ 
         & 8 & 6.65 & 0.1389 & 90.5 & 90.5 & 4.31 & 0.0997 & 67 & 65.76 \\ 
          & 12 & 6.01 & -0.0267 & 94.5 & 94.5 & 4.24 & 0.0022 & 66 & 64.99 \\
          \bottomrule
    \end{tabular}}
     \caption{The Inversion and Utility Performance on Classification Tasks on SST2 dataset with local DP. From a defender's perspective, $\uparrow$ means higher are better, $\downarrow$ means lower are better.}
     \label{tab:ldp_sst2}
\end{table*}

\begin{table*}[htbp]
    \centering
    \resizebox{0.65\linewidth}{!}{
    \begin{tabular}{l|l|cccc|cccc}
            \toprule

        Victim Model & $\epsilon$ & Rouge-L & COS & ACC & F1 & Rouge-L & COS & ACC & F1 \\ 
        \midrule
        ~ & &  \multicolumn{4}{c|}{LapMech} &  \multicolumn{4}{c}{PurMech} \\ 
         \midrule
          & - & 23.04 & 0.9219 & 71.5 & 70.37 & 23.04 & 0.9219 & 71.5 & 70.37 \\
          \cmidrule{2-10}
        \textsc{T5} & 1 & 4.23 & 0.0109 & 52.5 & 52.49 & 4.19 & -0.0235 & 55 & 54.93 \\ 
          & 4 & 4.23 & -0.0526 & 61.5 & 60.67 & 4.48 & 0.0322 & 63.5 & 63.39 \\ 
          & 8 & 4.19 & 0.0457 & 68.5 & 67.5 & 4.29 & -0.1135 & 71 & 70.03 \\ 
          & 12 & 4.33 & 0.0082 & 72 & 71.06 & 4.29 & -0.0210 & 71 & 70.03 \\ 
                 \midrule

          & - & 13.22 & 0.8585 & 70.5 & 68.45 & 13.22 & 0.8585 & 70.5 & 68.45 \\ 
          \cmidrule{2-10}
       \textsc{ GTR} & 1 & 4.54 & -0.0107 & 51 & 50.82 & 4.29 & -0.0079 & 50 & 49.82 \\ 
          & 4 & 4.89 & 0.0635 & 61.5 & 59.81 & 4.55 & -0.0066 & 61 & 58.4 \\ 
          & 8 & 4.55 & 0.0415 & 67 & 65.48 & 4.82 & -0.0179 & 67 & 65.16 \\ 
          & 12 & 4.64 & 0.0673 & 70.5 & 68.93 & 4.85 & 0.0439 & 70 & 68.75 \\
                 \midrule

          & - & 19.82 & 0.9212 & 68 & 66.8 & 19.82 & 0.9212 & 68 & 66.8 \\ 
          \cmidrule{2-10}
        \textsc{mT5} & 1 & 4.54 & -0.0650 & 47 & 46.99 & 4.52 & 0.0507 & 53 & 52.98 \\ 
          & 4 & 4.31 & 0.0432 & 61.5 & 60.67 & 4.14 & -0.0377 & 63.5 & 62.6 \\ 
          & 8 & 4.85 & 0.0386 & 65.5 & 64.86 & 4.31 & 0.0997 & 67 & 65.76 \\ 
          & 12 & 4.62 & 0.1038 & 66 & 64.99 & 4.24 & 0.0022 & 66 & 64.99 \\ 
                 \midrule
            & - & 17.88 & 0.9070 & 64.5 & 64.18 & 4.66 & -0.0340 & 62 & 61.69 \\ 
          \cmidrule{2-10}
        \textsc{mBERT} & 1 & 4.6 & 0.0153 & 54 & 54 & 17.88 & 0.9070 & 64.5 & 64.18 \\ 
          
         & 4 & 5 & 0.1196 & 59 & 58.9 & 4.23 & 0.0215 & 49.5 & 49.5 \\ 
          & 8 & 4.4 & 0.0156 & 60.5 & 60.28 & 4.6 & -0.0190 & 63.5 & 63.46 \\ 
          & 12 & 4.76 & 0.0117 & 61.5 & 61.45 & 4.26 & 0.0942 & 63 & 62.91 \\ 
        
                 \midrule

          & 0 & 17.13 & 0.9224 & 71.5 & 70.12 & 17.13 & 0.9224 & 71.5 & 70.12 \\ 
          \cmidrule{2-10}
        \shortstack{\textsc{OpenAI}\\ \textsc{(ada-2)}}   & 1 & 4.45 & 0.0041 & 53 & 52.83 & 3.89 & -0.0560 & 55 & 54.93 \\ 
          & 4 & 4.84 & 0.0572 & 63 & 62.46 & 4.09 & -0.0255 & 66.5 & 65.03 \\ 
         & 8 & 4.66 & -0.0191 & 68.5 & 66.97 & 4.9 & -0.0166 & 68.5 & 66.66 \\ 
        & 12 & 4.84 & 0.0728 & 71.5 & 69.83 & 5.3 & 0.1235 & 69.5 & 67.72 \\ 
                  \bottomrule

    \end{tabular}}
    \caption{The Inversion and Utility Performance on Classification Tasks on S140 dataset with local DP. From a defender's perspective, $\uparrow$ means higher are better, $\downarrow$ means lower are better.}
    \label{tab:s140_ldp}
\end{table*}

\end{document}